\documentclass[12pt]{JHEP}

\usepackage{epsfig}
\usepackage{cite}
\usepackage{latexsym}
\newcommand{\newsection}{    % Numeration of eqs. is automatic
\setcounter{equation}{0}
\section}
\renewcommand{\appendix}[1]{
    \addtocounter{section}{1}
    \setcounter{equation}{0}
    \renewcommand{\thesection}{\Alph{section}}
    \section*{Appendix \thesection\protect\indent #1}
    \addcontentsline{toc}{section}{Appendix \thesection\ \ \ #1}
}
\newcommand\encadremath[1]{\vbox{\hrule\hbox{\vrule\kern8pt
\vbox{\kern8pt \hbox{$\displaystyle #1$}\kern8pt}
\kern8pt\vrule}\hrule}}
\def\enca#1{\vbox{\hrule\hbox{
\vrule\kern8pt\vbox{\kern8pt \hbox{$\displaystyle #1$}
\kern8pt} \kern8pt\vrule}\hrule}}

\newcommand\figureframex[3]{
\begin{figure}[bth]
\hrule\hbox{\vrule\kern8pt
\vbox{\kern8pt \vbox{
\begin{center}
{\mbox{\epsfxsize=#1.truecm\epsfbox{#2}}}
\end{center}
\caption{#3}
}\kern8pt}
\kern8pt\vrule}\hrule
\end{figure}
}
\newcommand\figureframey[3]{
\begin{figure}[bth]
\hrule\hbox{\vrule\kern8pt
\vbox{\kern8pt \vbox{
\begin{center}
{\mbox{\epsfysize=#1.truecm\epsfbox{#2}}}
\end{center}
\caption{#3}
}\kern8pt}
\kern8pt\vrule}\hrule\end{figure}
}

\renewcommand{\thesection}{\arabic{section}}

\makeatletter
\@addtoreset{equation}{section}
\makeatother
\newtheorem{theorem}{Theorem}[section]

\newtheorem{remark}{Remark}[section]
\newtheorem{proposition}{Proposition}[section]
\newtheorem{lemma}{Lemma}[section]
\newtheorem{corollary}{Corollary}[section]
\newtheorem{definition}{Definition}[section]
\newcommand{\eq}[1]{eq.~(\ref{#1})}
\def\br{\begin{remark}\rm\small}
\def\er{\end{remark}}
\def\bt{\begin{theorem}}
\def\et{\end{theorem}}
\def\bd{\begin{definition}}
\def\ed{\end{definition}}
\def\bp{\begin{proposition}}
\def\ep{\end{proposition}}
\def\bl{\begin{lemma}}
\def\el{\end{lemma}}
\def\bc{\begin{corollary}}
\def\ec{\end{corollary}}
\def\beaq{\begin{eqnarray}}
\def\eeaq{\end{eqnarray}}
\newcommand{\proof}[1]{{\noindent \bf proof:}\par
{#1} $\bullet$}

\newcommand{\beq}{\begin{equation}}
\newcommand{\eeq}{\end{equation}}
\newcommand{\bea}{\begin{eqnarray}}
\newcommand{\eea}{\end{eqnarray}}

\newcommand\eol{\hspace*{\fill}\linebreak}
\newcommand\eop{\vspace*{\fill}\pagebreak}

\newcommand{\vs}{\vspace{0.7cm}}
\renewcommand{\and}{{\qquad {\rm and} \qquad}}

\newcommand{\virg}{{\qquad , \qquad}}

 \newcommand{\Tr}{{\,\rm Tr}\:}
\newcommand{\tr}{{\,\rm tr}\:}

\newcommand{\Res}{\mathop{\,\rm Res\,}}

\newcommand{\td}[1]{{\tilde{#1}}}

\newcommand{\e}{{\,\rm e}\,}
\newcommand{\ee}[1]{{{\rm e}^{#1}}}

\newcommand{\Pint}{{\int\kern -1.em -\kern-.25em}}

\newcommand{\Pol}{\mathop{\mathrm{Pol}}}

\newcommand{\Sym}{\Sigma}
\newcommand{\Symp}{{\ovl\Sigma}}

\newcommand{\ovl}{\overline}
\newcommand{\diag}{{\rm diag}}
\newcommand{\Cat}{{\rm Cat}\,}

\preprint{SPhT-T05/037, hep-th/0504029}

\title{Mixed correlation functions in the 2-matrix model, and the Bethe Ansatz}

\author{B.\ Eynard, N. \ Orantin \\
Service de Physique Th\'eorique de Saclay, CEA/DSM/SPhT,\\
Unit\'e de Recherche associ\'ee au CNRS (URA D2306), CEA Saclay,\\
F-91191 Gif-sur-Yvette Cedex, France.\\
 E-mail: eynard@spht.saclay.cea.fr, orantin@spht.saclay.cea.fr}

\abstract{Using loop equation technics, we compute all mixed traces correlation functions of the 2-matrix model to large N leading order.
The solution turns out to be a sort of Bethe Ansatz, i.e. all correlation functions can be decomposed on products of 2-point functions.
We also find that, when the correlation functions are written collectively as a matrix, the loop equations are equivalent to commutation relations.}


\begin{document}

%\topmargin .5cm \textheight 21.5cm \textwidth 15.8cm
%\oddsidemargin 0.54cm
%\evensidemargin 0.54cm
%\sloppy

%\maketitle

%%%%%%%%%%%%%%%%%%%%%%%%%%%%%%%%%%%%%%%%%%%%%%%%%%%%%%%%%%%%%%%%%%%%%%%%%%%%%%

%\pagestyle{empty}
%\hfill SPhT-T05/037
%\addtolength{\baselineskip}{0.20\baselineskip}
%\begin{center}
%\vspace{26pt}
%{\large \bf {Mixed correlation functions in the 2-matrix model, and the Bethe Ansatz}}
%\newline
%\vspace{26pt}

%{\sl B.\ Eynard}\hspace*{0.05cm}\footnote{ E-mail: eynard@cea.fr }
%,{\sl N.\ Orantin}\hspace*{0.05cm}\footnote{ E-mail: orantin@cea.fr }\\
%\vspace{6pt}
%Service de Physique Th\'{e}orique de Saclay,\\
%F-91191 Gif-sur-Yvette Cedex, France.\\
%\end{center}

%\vspace{20pt}
%\begin{center}
%{\bf Abstract}:

%\end{center}

%-----------------------------ABSTRACT--------------------------------------
%
%Abstract

%\begin{center}

%\end{center}

%\newpage
%\pagestyle{empty}

%\section*{}

%\newpage
%\vspace{26pt}
%\pagestyle{plain}
%\setcounter{page}{1}
%*********************************************************************
%==================== ARTICLE ========================================
%*********************************************************************

\newsection{Introduction}

Formal random matrix models have been used for their interpretation as combinatorial generating functions
for discretized surfaces \cite{Mehta, BIPZ, ZJDFG}.
The hermitean one-matrix model counts surfaces made of polygons of only one color, whereas  the hermitean
two--matrix model counts surfaces made of polygons of two colors.
In that respect, the 2-matrix model is more appropriate for the purpose of studying surfaces with non-uniform
boundary conditions.
At the continuum limit, the 2-matrix model gives access to ``boundary operators'' in conformal field theory \cite{kostov}.

Generating functions for surfaces with boundaries are obtained as random matrix expectation values.
The expectation value of a product of $l$ traces is the generating function for surfaces with $l$ boundaries,
the total power of matrices in each trace being the length of the corresponding boundary.
If each trace contains only one type of matrix (different traces may contain different types of matrices),
the expectation value is the generating function counting surfaces with uniform boundary conditions.
Those non-mixed expectation values have been computed for finite $n$ since the work of \cite{eynardmehta, Mehta2} and refined by \cite{Bergere1}.

Mixed correlation functions have been considered as a difficult problem for a long time and progress have been obtained only recently \cite{BEmixed, eynprats}.
Indeed, non-mixed expectation values can easily be written in terms of eigenvalues only (since the trace of a matrix is clearly related to its eigenvalues),
whereas  mixed correlation functions cannot ($Tr M_1^k M_2^{k'}$ cannot be written in terms of eigenvalues of $M_1$ and $M_2$).

The large $N$ limit of the generating function of the bicolored disc (i.e. one boundary, two colors, i.e. $<Tr M_1^k M_2^{k'}>$) has
been known since \cite{kaz, eynchain, eynchaint}.
The large $N$ limit of the generating function of the 4-colored disc (i.e. one boundary, 4 colors, i.e. $<Tr M_1^k M_2^{k'} M_1^{k''} M_2^{k'''}>$) has
been known since \cite{eynm2m}.
The all order expansion of correlation functions for the 1-matrix model has been obtained by a Feynman-graph representation in \cite{eynloop1mat}
and the generalization to non-mixed correlation functions of the 2-matrix model has been obtained in \cite{eoloop2mat}.

Recently, the method of integration over the unitary group of \cite{eynprats} has allowed to compute, for finite $N$, all mixed correlation functions of the 2-matrix model
in terms of orthogonal polynomials.

\smallskip

The question of computing mixed correlation functions in the large $N$ limit is addressed in the present article.

The answer is (not so) surprisingly related to classical results in integrable statistical models, i.e. the Bethe Ansatz.
It has been known for a long time that random matrix models are integrable in some sense (Toda, KP, KdV, isomonodromic systems,...),
but the relationship with Yang-Baxter equations and Bethe Ansatz was rather indirect.
The result presented in this article should give some new insight in that direction. We find that the $k$-point functions can be expressed
as the product of 2-point function, which is the underlying idea of the Bethe Ansatz.

\bigskip

{\noindent \bf Outline of the article:}

- section 1 is an introduction,

- in section 2, we set definitions of the model and correlation functions, and we write the relevant loop equations,

- in section 3, we introduce a Bethe Ansatz-like formula, and prove it in section~4,

- in section 5, we solve the problem under a matrix form,

- section 6 is dedicated to the special Gaussian case.

\newsection{The 2-matrix model, definitions and loop equations}

\subsection{Partition function}

We are interested in the formal matrix integral:
\beq\label{Zdef}
Z:=\int_{H_N^2} dM_1\, dM_2\, \ee{-N\Tr[V_1(M_1)+V_2(M_2)+M_1 M_2]}
\eeq
where $M_1$ and $M_2$ are $N\times N$ hermitean matrices and $dM_1$ (resp. $dM_2$) is the product of Lebesgue
measures of all independent real components of $M_1$ (resp. $M_2$).
$V_1(x)$ and $V_2(y)$ are complex polynomials of degree $d_1+1$ and $d_2+1$, called ``potentials''.
The formal matrix integral is defined as a formal power series in the coefficients of the potentials (see \cite{ZJDFG}),
computed by the usual Feynman method:
consider a local extremum of $\ee{-N\Tr[V_1(M_1)+V_2(M_2)+M_1 M_2]}$, and  expand the non quadratic part
as a power series and, for each term of the series, perform the Gaussian integration with the quadratic part.
This method does not care about the convergence of the integral, or of the series, it makes sense only order by order
and it is in that sense that it can be interpreted as the generating function of discrete surfaces.
All quantities in that model have a well defined $1/N^2$ expansion \cite{thoft}.

The extrema of $V_1(x)+V_2(y)+xy$ are such that:
\beq
V'_1(x)=-y \virg V'_2(y)=-x
\eeq
there are $d_1 d_2$ solutions (indeed $V'_2(-V'_1(x))=-x$), which we note $(\ovl{x}_I,\ovl{y}_I)$, \eol
$I=1,\dots, d_1 d_2$.
The extrema of $\Tr[V_1(M_1)+V_2(M_2)+M_1 M_2]$ can be chosen diagonal (up to a unitary transformation), with $\ovl{x}_I$'s and $\ovl{y}_I$'s on the diagonal:
\bea
M_1&=&\diag(
{\stackrel{n_1\,{\rm times}}{\overbrace{\ovl{x}_1,\dots,\ovl{x}_1}}},
{\stackrel{n_2\,{\rm times}}{\overbrace{\ovl{x}_2,\dots,\ovl{x}_2}}},
\dots
,{\stackrel{n_{d_1 d_2}\,{\rm times}}{\overbrace{\ovl{x}_{d_1 d_2},\dots,\ovl{x}_{d_1 d_2}}}})\cr
M_2&=&\diag(
{\stackrel{n_1\,{\rm times}}{\overbrace{\ovl{y}_1,\dots,\ovl{y}_1}}},
{\stackrel{n_2\,{\rm times}}{\overbrace{\ovl{y}_2,\dots,\ovl{y}_2}}},
\dots
,{\stackrel{n_{d_1 d_2}\,{\rm times}}{\overbrace{\ovl{y}_{d_1 d_2},\dots,\ovl{y}_{d_1 d_2}}}})
\eea
The extremum around which we perform the expansion is thus characterized by a set of filling fractions:
\beq
\epsilon_I = {n_I\over N}
\virg
\sum_{I=1}^{d_1 d_2} \epsilon_I=1
\eeq

To summarize, let us say that the formal matrix integral is defined for given potentials and filling fractions.

The ``one-cut'' case is the one where one of the filling fractions is $1$, and all the others vanish.
This is the case where the Feynman expansion is performed in the vicinity of only one extremum.

\subsection{Enumeration of discrete surfaces}

It is well known that formal matrix integrals are generating functions for the enumeration of discrete surfaces \cite{ZJDFG, BIPZ, Kazakov, courseynard}.

\smallskip

For instance, in the one-cut case (expansion near an extremum $\ovl{x},\ovl{y}$), one has:
\beq
\begin{array}{l}
-\ln{Z} =\cr
\sum_{G} {1\over \#{\rm Aut}(G)} N^{\chi(G)} \left({g_2\over \delta}\right)^{n_{--}(G)}\left({\td{g}_2\over \delta}\right)^{n_{++}(G)}\left({-1\over \delta}\right)^{n_{+-}(G)}
\prod_{i=3}^{d_1+1} g_i^{n_i(G)} \prod_{i=3}^{d_2+1} \td{g}_i^{\td{n}_i(G)}\cr
\end{array}
\eeq
where the summation is over all finite connected closed discrete surfaces made  of polygons of two signs (+ and -).
For such a surface (or graph) $G$, $\chi(G)$ is its Euler characteristic, $n_i(G)$ is the number of $i$-gons carrying a $+$ sign, $\td{n}_i(G)$ is the number of $i$-gons carrying a $-$ sign,
$n_{++}(G)$ is the number of edges separating two $+$ polygons,
$n_{--}(G)$ is the number of edges separating two $-$ polygons and $n_{+-}(G)$ is the number of edges separating two polygons of different signs.
$\#{\rm Aut}(G)$ is the number of automorphisms of $G$.

The $g_i$'s, $\td{g}_i$'s and $\delta$ are defined as follows:
\beq
g_k := \left.{\partial^k V_1(x)\over \partial x^k}\right|_{x=\ovl{x}}
\virg
\td{g}_k := \left.{\partial^k V_2(y)\over \partial x^k}\right|_{x=\ovl{y}}
\virg
\delta := g_2 \td{g}_2-1
\eeq

Example of a discrete surface:
\beq
\begin{array}{r}
{\epsfysize 6cm\epsffile{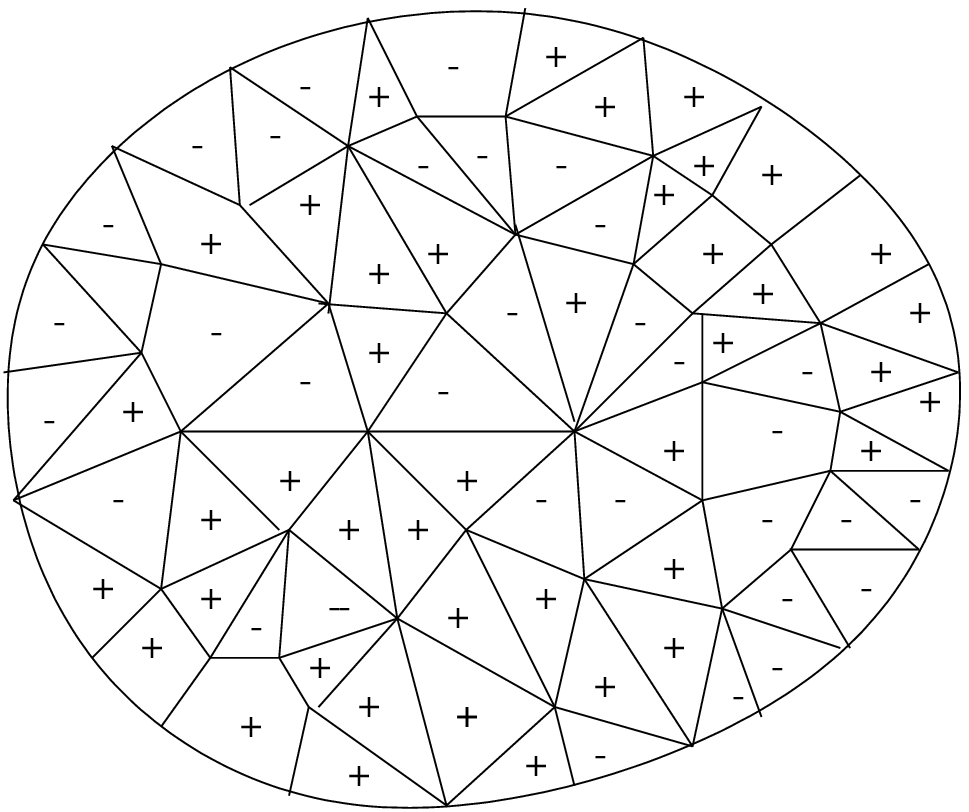}}
\end{array}
\eeq

In the multicut case, i.e.  with arbitrary filling fractions, matrix integrals can still be interpreted in terms of ``foams'' of surfaces,
and we refer the reader to the appendix of \cite{BDE} or to \cite{eynhabilit} for more details.

\subsection{Enumeration of discrete surfaces with boundaries}

Similarly, given a sequence of signs $s_1,s_2,\dots, s_k$, $s_i\in{1,2}$, it is well known that the following quantity:
\beq\label{Trdisc}
\left<\Tr(\prod_{i=1}^{k} M_{s_i})\right>
\eeq
is the generating function of discrete surfaces with one boundary of length $k$, whose signs of polygons on the edges are given by the sequence $(s_1,\dots,s_k)$.

\medskip
Example of a discrete surface with boundary $(++++++-----+++++------)$:
\beq
\begin{array}{lll}
\left<\Tr(M_1^{6}M_2^{5}M_1^{5}M_2^{6})\right> \,\,& =& \,\,\, \sum_G\,\,\, {\epsfysize 5cm\epsffile{surfdiscr.eps}}
\end{array}
\eeq

More generally, an expectation value of a product of $n$ traces is the generating function for discrete surfaces with $n$ boundaries.

In this article, we are interested only in one boundary and to leading order in $N$, i.e. surfaces with the topology of a disc.

\subsection{Master loop equation and algebraic curve}

Let us define:
\beq
W(x):={1\over N}\left<\Tr{1\over x-M_1}\right>
\virg
\td{W}(y):={1\over N}\left<\Tr{1\over y-M_2}\right>
\eeq
where the expectation values are formally computed as explained in the previous section, with the weight
$\ee{-N\Tr[V_1(M_1)+V_2(M_2)+M_1 M_2]}$.
$W(x)$ (resp. $\td{W}(y)$) is defined as a formal power series in its large $x$ (resp. large $y$) expansion,
as well as in the expansion in the coefficients of the potentials.
$W(x)$ (resp. $\td{W}(y)$) is a generating function for surfaces with one uniform boundary, i.e. with only sign $+$ (resp. sign $-$)
polygons touching the boundary by an edge:
\beq
\begin{array}{r}
W(x)={\epsfysize 2cm\epsffile{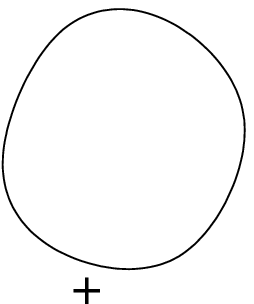}}
\virg
\td{W}(y)={\epsfysize 2cm\epsffile{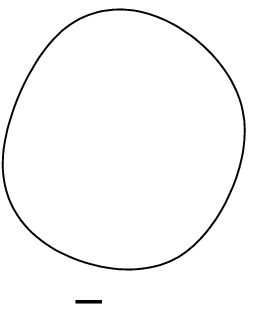}}
\end{array}
\eeq

We also define the following formal series:
\beq
Y(x):=W(x)-V'_1(x)
\virg
X(y):=\td{W}(y)-V'_2(y)
\eeq

In addition, we define:
\beq
P(x,y):={1\over N}\left<\Tr{V'_1(x)-V'_1(M_1)\over x-M_1}{V'_2(y)-V'_2(M_2)\over y-M_2}\right>
\eeq
\beq
U(x,y):={1\over N}\left<\Tr{1\over x-M_1}{V'_2(y)-V'_2(M_2)\over y-M_2}\right>+x+V'_2(y)
\eeq
\bea
U(x,y;x')&:=&\left<\Tr{1\over x-M_1}{V'_2(y)-V'_2(M_2)\over y-M_2}\Tr{1\over x'-M_1}\right>\cr
&& - N^2 W(x') (U(x,y)-x-V'_2(y))\cr
\eea
\beq
E(x,y):=(V'_1(x)+y)(V'_2(y)+x)+P(x,y)-1
\eeq
Notice that $U(x,y)$ and $U(x,y;x')$ are polynomials of $y$ (with degree at most $d_2-1$),
 $P(x,y)$ is a polynomial of both variables of degree ($d_1-1,d_2-1$) and $E(x,y)$ is a polynomial
of both $x$ and $y$ of degree $(d_1+1,d_2+1)$.

It has been obtained in many articles \cite{eynmultimat, staudacher, eynchain, eynchaint}, that:
\beq
E(x,Y(x))={1\over N^2} U(x,Y(x),x)
\eeq
To large $N$ leading order that equation reduces to an algebraic equation for $Y(x)$, called the ``Master loop equation'' \cite{staudacher}:
\beq
E(x,Y(x))=0
\eeq
(similarly, one also has $E(X(y),y)=0$, which implies $Y\circ X={\rm Id}$, known as Matytsin's equation \cite{matytsin}).
The coefficients of $E(x,y)$, i.e. of $P(x,y)$, are entirely determined by the conditions $\oint_{{\cal A}_i} ydx = 2i\pi \epsilon_i$
for a choice of irreducible cycles on the algebraic curve.

The properties of that algebraic equation have been studied in many works \cite{eynmultimat, KazMar}. Here we assume that it is known.

\subsection{Correlation functions, definitions}

We define:
\beq
 \ovl{W}_k(x_1,y_1,x_2,\dots,x_k,y_k)
:={1\over N}\left<\Tr\prod_{j=1}^k {1\over x_j-M_1}{1\over y_j-M_2}\right>
\eeq
\bea
&& \ovl{U}_k(x_1,y_1,x_2,\dots,x_k,y_k) \cr
&:=& \Pol_{y_k} V'_2(y_k)\, \ovl{W}_k(x_1,y_1,x_2,\dots,x_k,y_k) \cr
&=&{1\over N}\left<\Tr {1\over x_1-M_1}{1\over y_1-M_2}\,\dots \,{1\over x_{k}-M_1}{V'_2(y_k)-V'_2(M_2)\over y_k-M_2}\right> \cr
\eea
\bea
&& \ovl{P}_k(x_1,y_1,x_2,\dots,x_k,y_k) \cr
&:=& \Pol_{x_1} \Pol_{y_k}  V'_1(x_1)\, V'_2(y_k)\, \ovl{W}_k(x_1,y_1,x_2,\dots,x_k,y_k)  \cr
&=&{1\over N}\left<\Tr {V'_1(x_1)-V'_1(M_1)\over x_1-M_1}\,{1\over y_1-M_2}\dots{1\over x_k-M_1}\,\,{V'_2(y_k)-V'_2(M_2)\over y_k-M_2}\right> \cr
\eea
\beq
A_k(x_1,y_1,x_2,\dots,x_k):={1\over N}\left<\Tr {1\over x_1-M_1}{1\over y_1-M_2}\dots{1\over x_k-M_1}V'_2(M_2)\right>
\eeq

where $\Pol_x f(x)$ denotes the polynomial part at infinity of $f(x)$ (i.e. the positive part in the Laurent series for $x$ near infinity).

The functions $\ovl{W}_k$ are generating functions for discrete discs with all possible boundary conditions.
One can recover any generating function of type \eq{Trdisc} by expanding into powers of the $x_i$'s and $y_i$'s.

\medskip

For convenience, we prefer to consider the following functions:
\beq
W_k(x_1,y_1,x_2,\dots,x_k,y_k):=\ovl{W}_k(x_1,y_1,x_2,\dots,x_k,y_k)+\delta_{k,1}
\eeq
\beq
U_k(x_1,y_1,x_2,\dots,x_k,y_k):=\ovl{U}_k(x_1,y_1,x_2,\dots,x_k,y_k)+\delta_{k,1}(V'_2(y_k)+x_k)
\eeq
and for $k>1$:
\beq
P_k(x_1,y_1,x_2,\dots,x_k,y_k):=\ovl{P}_k(x_1,y_1,x_2,\dots,x_k,y_k)+{W}_{k-1}(x_{2},\dots,x_k,y_1)
\eeq

For the smallest values of $k$, those expectation values can be found in the literature to large $N$ leading order:

$\bullet$ it was found in \cite{eynchain,eynchaint, eynmultimat}:
\beq\label{W1U1}
W_1(x,y) = {E(x,y)\over (x-X(y))(y-Y(x))}
\virg
U_1(x,y) = {E(x,y)\over (y-Y(x))}
\eeq

$\bullet$ it was found in the appendix C of \cite{eynm2m} (there is a change of sign, because the action in \cite{eynm2m} was $\e^{-N\tr(V_1(M_1)+V_2(M_2)-M_1M_2)}$):
\beq
W_2(x_1,y_1,x_2,y_2) = {W_1(x_1,y_1)W_1(x_2,y_2)-W_1(x_1,y_2)W_1(x_2,y_1)\over (x_1-x_2)(y_1-y_2)}
\eeq

$\bullet$ For finite $N$, it was found in \cite{BEmixed}, and with notations explained in \cite{BEmixed}:
\beq\label{BEmixedW1}
W_1(x,y) = \det{\left(1_N+\Pi_{N-1}{1\over x-Q}{1\over y-P^t}\Pi_{N-1}\right)}
\eeq

$\bullet$ For finite $N$, it was found in \cite{eynprats} how to compute any mixed correlation function in terms of determinants involving biorthogonal polynomials, with a formula very similar to \eq{BEmixedW1}.

\medskip

Here, we shall find a formula for all $W_k$'s in the large $N$ limit.

\subsection{Loop equations}

Loop equations are nothing but Schwinger--Dyson equations.
They are obtained by writing that an integral is invariant under a change of variable,
or alternatively by writing that the integral of a total derivative vanishes.

The loop equation method is well known and explained in many works \cite{eynmultimat, staudacher}.
Here, we write for each change of variable the corresponding loop equation (we use a presentation similar to that of \cite{eynmultimat}).

In all what follows we consider $k>1$.

\bigskip

$\bullet$ the change of variable: $\delta M_2={1\over x_1-M_1}{1\over y_1-M_2}\,\dots \,{1\over x_{k}-M_1}$ implies:
\bea\label{loopeqA}
A_k(x_1,\dots,x_k)
&=& \sum_{j=1}^{k-1} \ovl{W}_j(x_1,\dots,y_j)\,\ovl{W}_{k-j}(x_k,y_j,\dots,y_{k-1}) \cr
&& + { x_1\ovl{W}_{k-1}(x_{1},y_1,\dots,y_{k-1})- x_k\ovl{W}_{k-1}(x_{k},y_1,\dots,y_{k-1})\over x_1-x_k} \cr
&=& \sum_{j=1}^{k-1} {W}_j(x_1,\dots,y_j)\,{W}_{k-j}(x_k,y_j,\dots,y_{k-1}) \cr
&& - {W}_{k-1}(x_k,y_1,\dots,y_{k-1})  \cr
&& + x_k\,{ {W}_{k-1}(x_{1},y_1,\dots,y_{k-1})- {W}_{k-1}(x_{k},y_1,\dots,y_{k-1})\over x_1-x_k} \cr
\eea

$\bullet$ the change of variable: $\delta M_1={1\over x_1-M_1}{1\over y_1-M_2}\,\dots \,{1\over x_{k}-M_1}{V'_2(y_k)-V'_2(M_2)\over y_{k}-M_2}$ implies:
\beq\label{loopeqU}
\begin{array}{l}
(Y(x_1)-y_k)\, \ovl{U}_k(x_1,\dots,y_k) \cr
= \sum_{j=2}^k {W_{j-1}(x_1,y_1,\dots,y_{j-1})-W_{j-1}(x_j,y_1,\dots,y_{j-1})\over x_1-x_j}\,\ovl{U}_{k-j+1}(x_j,y_j,\dots,x_k,y_k) \cr
+ V'_2(y_k) {W_{k-1}(x_1,y_1,\dots,y_{k-1})-W_{k-1}(x_k,y_1,x_2,\dots,y_{k-1})\over x_1-x_k} \cr
+ A_k(x_1,\dots,x_k) - \ovl{P}_k(x_1,y_1,x_2,\dots,x_k,y_k) \cr
= \sum_{j=2}^k {W_{j-1}(x_1,y_1,\dots,y_{j-1})-W_{j-1}(x_j,y_1,\dots,y_{j-1})\over x_1-x_j}\,{U}_{k-j+1}(x_j,y_j,\dots,x_k,y_k) \cr
+\sum_{j=1}^{k-1} {W}_j(x_1,\dots,y_j)\,{W}_{k-j}(x_k,y_j,\dots,y_{k-1}) \cr
- {P}_k(x_1,y_1,x_2,\dots,x_k,y_k) \cr
\end{array}
\eeq
where we have used the loop equation \eq{loopeqA} for $A_k(x_1,\dots,x_k)$.

$\bullet$ the change of variable: $\delta M_2={1\over x_1-M_1}{1\over y_1-M_2}\,\dots \,{1\over x_{k}-M_1}{1\over y_{k}-M_2}$ implies:
\bea\label{loopeqW}
&& (X(y_k)-x_1)\,\ovl{W}_k(x_1,y_1,x_2,\dots,x_k,y_k) \cr
&=& \sum_{j=1}^{k-1} {{W}_{k-j}(x_{j+1},\dots,y_k)-{W}_{k-j}(x_{j+1},\dots,x_k,y_j)\over y_k-y_j}\, {W}_j(x_1,\dots,y_j) \cr
&& - U_k(x_1,\dots,y_k)   \cr
\eea

\subsection{Recursive determination of the correlation functions}

\bt\label{thloopdetermineWk}
The system of equations \eq{loopeqU} and \eq{loopeqW} for all $k$ has a unique solution.
\et

In other words, if we can find some functions $W_k$, $U_k$, $P_k$ which obey \eq{loopeqU} and \eq{loopeqW}
for all $k$, then they are the correlation functions we are seeking.

\medskip
\proof{
$W_1$, $U_1$ and $P_1$ have already been computed in the literature.

Assume that we have computed $W_j$, $U_j$, $P_j$ for all $j<k$.
Let us show that \eq{loopeqU} and \eq{loopeqW} determine uniquely $W_k$, $U_k$ and $P_k$.

Let $X^{(\alpha)}(y_k)$, $\alpha=0,\dots,d_1$ be the $d_1+1$ solutions for $x$ of $E(x,y_k)=0$.
For every $\alpha=0,\dots, d_1$ one has:
\beq
Y(X^{(\alpha)}(y_k))=y_k
\eeq
At $x_1=X^{(\alpha)}(y_k)$, \eq{loopeqU} reads:
\beq\label{loopeqP}
\begin{array}{l}
{P}_k(X^{(\alpha)}(y_k),y_1,x_2,\dots,x_k,y_k) \cr
= \sum_{j=2}^k {W_{j-1}(X^{(\alpha)}(y_k),y_1,\dots,y_{j-1})-W_{j-1}(x_j,y_1,x_2,\dots,y_{j-1})\over X^{(\alpha)}(y_k)-x_j}\,{U}_{k-j+1}(x_j,y_j,\dots,x_k,y_k) \cr
+\sum_{j=1}^{k-1} {W}_j(X^{(\alpha)}(y_k),\dots,y_j)\,{W}_{k-j}(x_k,y_j,\dots,y_{k-1}) \cr
\end{array}
\eeq
where all the quantities in the RHS are known from the recursion hypothesis.
We thus know the value of $P_k$ for $d_1+1$ values of $x_1$. Since $P_k$ is a polynomial in $x_1$ of degree at most $d_1-1$,
we can determine $P_k$ by the interpolation formula:
\bea\label{loopeqinterpolP}
&& (x_1-X(y_k)) {{P}_k(x_1,\dots,y_k)\over E(x_1,y_k)}  \cr
&=& \sum_{\alpha=1}^{d_2} {(X^{(\alpha)}(y_k)-X(y_k))\,{P}_k(X^{(\alpha)}(y_k),\dots,y_k)\over (x_1-X^{(\alpha)}(y_k))\,E_x(X^{(\alpha)}(y_k),y_k)} \cr
\eea
where $X_k=X(y_k)$ denotes $X^{(0)}(y_k)$.
Once $P_k$ is known, \eq{loopeqU} allows to compute $U_k$, and \eq{loopeqW} allows to compute $W_k$.
}

\newsection{A Bethe Ansatz-like formula for correlation functions}

Thus, the loop equations determine $W_k$ uniquely,
i.e., if we can find $W_k$, $U_k=\Pol_{y_k} V'_2(y_k) W_k$ and $P_k=W_{k-1}+\Pol_{x_1} V'_1(x_1) U_k$ which satisfy \eq{loopeqW},\eq{loopeqU},
it means that we have the right solution.
We can thus make an Ansatz for $W_k$, and check that it satisfies the loop equations above.

\medskip

Our Ansatz is similar to the Bethe Ansatz \cite{gaudin}:
\beq\label{Ansatz}
W_k(x_1,y_1,\dots,x_k,y_k) = \sum_{\sigma\in \Symp_k}\, C^{(k)}_\sigma(x_1,y_1,\dots,x_k,y_k)\,\, \prod_{i=1}^k W_1(x_i,y_{\sigma(i)})
\eeq
where the coefficients $C_\sigma$ are rational fractions of the $x_i$'s and $y_i$'s, with at most simple poles at coinciding points, and
independent of the potentials.

We call \eq{Ansatz} a Bethe Ansatz-like formula, because it is very similar to the solution initially found by Bethe for the 1-dimensional spin chain, and then for
the $\delta$-interacting bosons.

\smallskip

If we assume that \eq{Ansatz} satisfies \eq{loopeqW}, we can in particular take the residue of \eq{loopeqW} at $y_k\to Y(x_l)$ for some $l$.
That implies the following relationship among the coefficients $C^{(k)}_\sigma$'s:
\bea\label{Crec}
&& (x_{\sigma^{-1}(k)}-x_1)\,C^{(k)}_\sigma(x_1,y_1,\dots, x_k,y_k) \cr
&=&
\sum_{j=1}^{k-1}
\sum_{\tau\in \Sym(1,\dots,j)}
\sum_{\rho\in \Sym(j+1,\dots,k)}\,
\delta_{\sigma,\tau\rho}\,\,
{C^{(j)}_\tau(x_1,\dots,y_j) C^{(k-j)}_\rho(x_{j+1},\dots,y_k)\over y_k-y_j}  \cr
\eea
Beside, since $W_k$ is the expectation value of a trace, the $C_\sigma^{(k)}$'s must be cyclically invariant:
\beq\label{Ccyclic1}
C^{(k)}_\sigma(x_1,y_1,x_2,y_2,\dots, x_k,y_k) = C^{(k)}_\sigma(x_2,y_2,\dots, x_k,y_k,x_1,y_1)
\eeq
and, since $W_k$ should have no poles at coinciding points $y_k=y_j$ one should have:
\beq\label{cancelpoleC1}
\Res_{y'_k\to y_j} C^{(k)}_\sigma(x_1,y_1,x_2,y_2,\dots, x_k,y'_k) dy'_k
=
\Res_{y'_k\to y_j} C^{(k)}_{(k,j)\circ\sigma}(x_1,y_1,x_2,y_2,\dots, x_k,y'_k) dy'_k
\eeq
With $C^{(1)}=1$, it is clear that the set of equations \eq{Crec}, \eq{Ccyclic1}, \eq{cancelpoleC1}, have at most a unique solution.
We prove in the next section that the solution exists, and thus, \eq{Crec}, \eq{Ccyclic1}, \eq{cancelpoleC1} determine $C^{(k)}_\sigma$ uniquely.
The $C^{(k)}_\sigma$'s are explicitly computed in section \ref{sectionCsigma}.

Then in section \ref{proofcorrel}, we prove that:

\bt\label{thansatzloop}
If the $C^{(k)}_\sigma$'s are rational functions defined by \eq{Crec}, \eq{Ccyclic1}, \eq{cancelpoleC1}, then the functions $\widehat{W}_k$'s defined by the RHS of  \eq{Ansatz},
the functions $\widehat{U}_k(x_1,\dots,y_k):=\Pol (V'_2(y_k))\widehat{W}_k(x_1,\dots,y_k)$,
and the functions $\widehat{P}_k(x_1,\dots,y_k):=\Pol (V'_1(x_1))\widehat{U}_k(x_1,\dots,y_k) + \widehat{W}_{k-1}(x_2,\dots,x_k,y_1)$,
satisfy \eq{loopeqW} and \eq{loopeqU}.
\et
As a corollary, using theorem \ref{thloopdetermineWk},  we have:
\bt\label{mainth}
\beq\label{thansatz}
\encadremath{
W_k(x_1,y_1,\dots,x_k,y_k) = \sum_{\sigma\in \Symp_k}\, C^{(k)}_\sigma(x_1,y_1,\dots,x_k,y_k)\,\, \prod_{i=1}^k W_1(x_i,y_{\sigma(i)})
}\eeq
\et

The derivation of theorem \ref{thansatzloop} is quite technical and is presented in section \ref{proofcorrel}.

\newsection{Amplitudes of permutations}
\label{sectionCsigma}

In this section, we compute the amplitudes $C^{(k)}_\sigma$ explicitly.

\smallskip

Eq. (\ref{Crec}), \eq{Ccyclic1}, \eq{cancelpoleC1} and initial condition $C^{(1)}(x_1,y_1)=1$ clearly define at most a unique function $C^{(k)}_\sigma(x_1,\dots,y_k)$.
In this section, we build the solution explicitly, and then, we prove that the function we have constructed indeed satisfies \eq{Crec}, \eq{Ccyclic1}, \eq{cancelpoleC1}.

Below we prove that \eq{Crec} implies that $C^{(k)}_\sigma$ vanishes for non planar permutations (see Definition 4.1 below)
and, for planar permutations, $C^{(k)}_\sigma$ is the product of $C^{(k)}_{\rm Id}$ corresponding to faces.
We are thus led to introduce the following definitions:

\subsection{Some definitions: planar permutations}

Let $S$ be the shift permutation:
\beq\label{desSshift}
S:={\rm shift} = (1,2, \dots, k-1,k) \virg {\rm i.e.}\quad S(i)=i+1
\eeq

\bd\label{Defplanar}
A permutation $\sigma\in \Sigma_k$ is called planar if
\beq\label{defplanarcycles}
n_{\rm cycles}(\sigma)+n_{\rm cycles}(S\sigma)=k+1
\eeq
where $n_{\rm cycles}(\sigma)$ is the number of irreducible cycles composing the permutation $\sigma$.

Let $\Symp_k\subset \Sigma_k$ be the set of planar permutations of rank $k$.

Eq. (\ref{defplanarcycles}) is equivalent to saying that if one draws the points $x_1,y_1,\dots,x_k,y_k$ on a circle, and draws a line between each pair $(x_j,y_{\sigma(j)})$,
the lines don't intersect. The cycles of $\sigma$ and the cycles of $S\sigma$ correspond to the faces (i.e. the connected components) of that partition of the disc.
\ed
Each planar permutation can also be represented as an arch system, and thus, the number of possible planar permutations
is given by the Catalan number $\Cat(k)$:
\beq
{\rm Card}(\Symp_k) = \Cat(k) = {2k!\over k! \,(k+1)!}
\eeq
\beq
\begin{array}{r}
{\epsfysize 6cm\epsffile{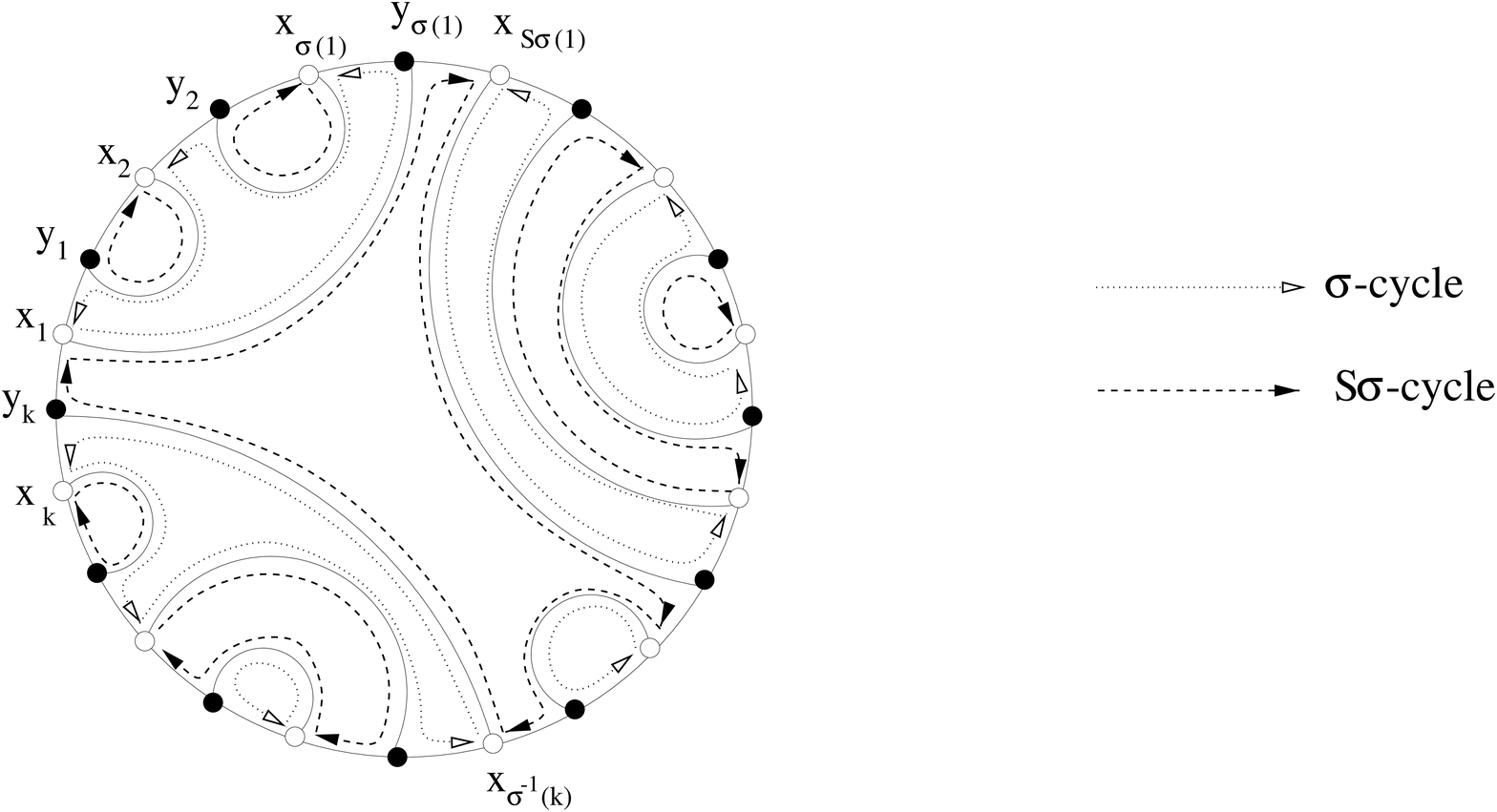}}
\end{array}
\eeq
\begin{center}
{\footnotesize
Example of a planar permutation and its faces.}
\end{center}

\subsection{Face amplitudes}

\bd\label{DefFsigma}
For any $k\geq 1$, we define a rational function of $x_1,\dots,y_k$:
\beq
F^{(k)}(x_1,y_1,x_2,\dots,x_k,y_k)
\eeq
by the recursion formula:
\bea\label{recF}
F^{(1)}(x_1,y_1) &:=& 1 \cr
F^{(k)}(x_1,y_1,\dots,x_k,y_k) &:=&  \sum_{j=1}^{k-1} {F^{(j)}(x_1,y_1,\dots,x_j,y_j) F^{(k-j)}(x_{j+1},y_{j+1},\dots,x_k,y_k) \over (x_k-x_1)(y_k-y_j)} \cr
\eea
\ed

\bl\label{lemFhomo}
For any $k\geq 1$ :
\beq
F^{(k)}(x_1,y_1,\dots,x_k,y_k) = O(y_k^{1-k})
\eeq
when $y_k \rightarrow \infty$.
\el
\proof{
Let us prove it by induction on k.
It is true for k=1. Let k be larger or equal to 2 and assume that this is true for all $F^{(j)}$ with $j<k$.
Then \eq{recF} straightforwardly gives the same behaviour for $F^{(k)}$.}

\bl\label{lemFcyclic}
$F^{(k)}$ has cyclic invariance, i.e.
\beq
F^{(k)}(x_2,y_2,\dots,x_k,y_k,x_1,y_1) = F^{(k)}(x_1,y_1,\dots,x_k,y_k)
\eeq
\el
\proof{
We prove it by recursion. It is clearly true for $k=1$ and $k=2$ since \eol
$F^{(2)}(x_1,y_1,x_2,y_2) = {1\over (x_2-x_1)(y_2-y_1)}$.
For $k\geq 3$, assume that it is true for all $F^{(j)}$ with $j<k$.
One has:
\bea
&& F^{(k)}(x_2,y_2,\dots,x_k,y_k,x_1,y_1) \cr
&=& \sum_{j=2}^{k} {F^{(j-1)}(x_2,\dots,x_j,y_j) F^{(k+1-j)}(x_{j+1},y_{j+1},\dots,x_k,y_k,x_1,y_1) \over (x_1-x_2)(y_1-y_j)} \cr
&=& \sum_{j=2}^{k} {F^{(j-1)}(x_2,\dots,x_j,y_j) F^{(k+1-j)}(x_1,y_1,x_{j+1},y_{j+1},\dots,x_k,y_k) \over (x_1-x_2)(y_1-y_j)} \cr
&=& \sum_{j=2}^{k} {F^{(j-1)}(x_2,\dots,x_j,y_j)\over (x_1-x_2)(y_1-y_j)} \cr
&& \qquad \sum_{l=j+1}^{k-1} {F^{(l+1-j)}(x_1,y_1,x_{j+1},y_{j+1},\dots,x_l,y_l) F^{(k-l)}(x_{l+1},y_{l+1},\dots,x_k,y_k)
\over (x_k-x_1)(y_k-y_l)} \cr
&& +\sum_{j=2}^{k} {F^{(j-1)}(x_2,\dots,x_j,y_j)\over (x_1-x_2)(y_1-y_j)}{F^{(1)}(x_1,y_1) F^{(k-j)}(x_{j+1},y_{j+1},\dots,x_k,y_k)
\over (x_k-x_1)(y_k-y_1)} \cr
&=& \sum_{l=3}^{k-1} {F^{(k-l)}(x_{l+1},y_{l+1},\dots,x_k,y_k)\over (x_k-x_1)(y_k-y_l)} \cr
&& \qquad  \sum_{j=2}^{l}{F^{(j-1)}(x_2,\dots,x_j,y_j)F^{(l+1-j)}(x_{j+1},y_{j+1},\dots,x_l,y_l,x_1,y_1)
\over (x_1-x_2)(y_1-y_j)} \cr
&=& \sum_{l=3}^{k-1} {F^{(k-l)}(x_{l+1},y_{l+1},\dots,x_k,y_k)F^{(l)}(x_2,\dots,x_l,y_l,x_1,y_1)\over (x_k-x_1)(y_k-y_l)} \cr
&=& \sum_{l=3}^{k-1} {F^{(l)}(x_1,y_1,x_2,\dots,x_l,y_l)F^{(k-l)}(x_{l+1},y_{l+1},\dots,x_k,y_k)\over (x_k-x_1)(y_k-y_l)} \cr
&=& F^{(k)}(x_1,y_1,\dots,x_k,y_k) \cr
\eea
}

\bl\label{lemFsimplepoles}
For $k\geq 2$, $F^{(k)}$ has simple poles in $y_k$:
\bea\label{49}
F^{(k)}(x_1,y_1,\dots,x_k,y_k) = \sum_{l=1}^{k-1} {1\over y_k-y_l}\,\Res_{y'_k\to y_l} F^{(k)}(x_1,y_1,\dots,x_k,y'_k) dy'_k
\eea
\el
\proof{It is clearly true for $k=2$. We prove it by induction on $k$.
Assume that it is true up to $k-1$.
Using the recursion hypothesis, one can see that each term in the RHS of \eq{recF} has at most a simple pole at $y_k=y_l$
and one can write \eq{49} with the use of Lemma \ref{lemFhomo}. Thus the
recursion hypothesis is true for $k$.
}

\subsection{The amplitudes $C_\sigma$}

\bd\label{defC}
Then, for any $k\geq 1$, and for any permutation $\sigma\in \Sigma_k$,
we define $C^{(k)}_\sigma(x_1,y_1,x_2,\dots,x_k,y_k)$ a rational function of $x_1,\dots,y_k$,
by:

$\bullet$
$\quad C^{(k)}_\sigma(x_1,y_1,x_2,\dots,x_k,y_k) := 0 $ if $\sigma$  is not planar, and

$\bullet$
if $\sigma$ is planar, we decompose $\sigma$ and $S\sigma$ into their product of cycles:
\beq
\sigma=\sigma_1 \sigma_2 \dots \sigma_l
\virg
S\sigma=\td\sigma_1 \td\sigma_2 \dots \td\sigma_{\td{l}}
\eeq
such that:
\beq
\sigma_j = (i_{j,1},i_{j,2},\dots,i_{j,l_j}) \virg \sigma(i_{j,m})=i_{j,m+1}
\eeq
\beq
\td\sigma_j = (\td{i}_{j,1},\td{i}_{j,2},\dots,\td{i}_{j,\td{l}_j}) \virg \sigma(\td{i}_{j,m})=\td{i}_{j,m+1}-1
\eeq

\bea\label{defCprodF}
C^{(k)}_\sigma(x_1,y_1,x_2,\dots,x_k,y_k)
&:=&\prod_{j=1}^l F^{(l_j)}(x_{i_{j,1}},y_{i_{j,2}},x_{i_{j,2}},y_{i_{j,3}},\dots,x_{i_{j,l_j}},y_{i_{j,1}}) \cr
&& \prod_{j=1}^{\td{l}} F^{(\td{l}_j)}(x_{\td{i}_{j,1}},y_{\td{i}_{j,2}-1},x_{\td{i}_{j,2}},\dots,y_{\td{i}_{j,\td{l}_j}-1},x_{\td{i}_{j,\td{l}_j}},y_{\td{i}_{j,1}-1}) \cr
\eea
i.e.
$C^{(k)}_\sigma$ is the product of $F$'s of each connected component of the disc partitioned by $\sigma$.
\ed

\subsection{Examples}

In particular with $\sigma=Id$, we have:
\beq
 C^{(k)}_{Id}(x_1,y_1,x_2,\dots,x_k,y_k)
= F^{(k)}(x_1,y_1,x_2,y_2,\dots,x_k,y_k)
\eeq
and with $\sigma=S^{-1}$:
\beq
 C^{(k)}_{S^{-1}}(x_1,y_1,x_2,\dots,x_k,y_k)
= F^{(k)}(x_k,y_{k-1},x_{k-1},\dots, y_1,x_1,y_k)
\eeq

\subsubsection*{An example with $k=12$}
Let us consider an example of $\sigma \in \overline{\Sigma}_{12}$ defined as follow:
$\sigma(1)=3$, $\sigma(2)=1$, $\sigma(3)=2$, $\sigma(4)=7$, $\sigma(5)=6$, $\sigma(6)=5$, $\sigma(7)=4$,
$\sigma(8)=8$, $\sigma(9)=11$, $\sigma(10)=10$, $\sigma(11)=12$ and $\sigma(12)=9$:
\bea
\sigma = (1,3,2)(4,7)(5,6)(8)(9,12,11)(10) \cr
S \sigma = (1,4,8,9)(2)(3)(5,7)(6)(10,11)(12)
\eea

The corresponding arch system is:
\beq\label{system}
\begin{array}{r}
{\epsfysize 6cm\epsffile{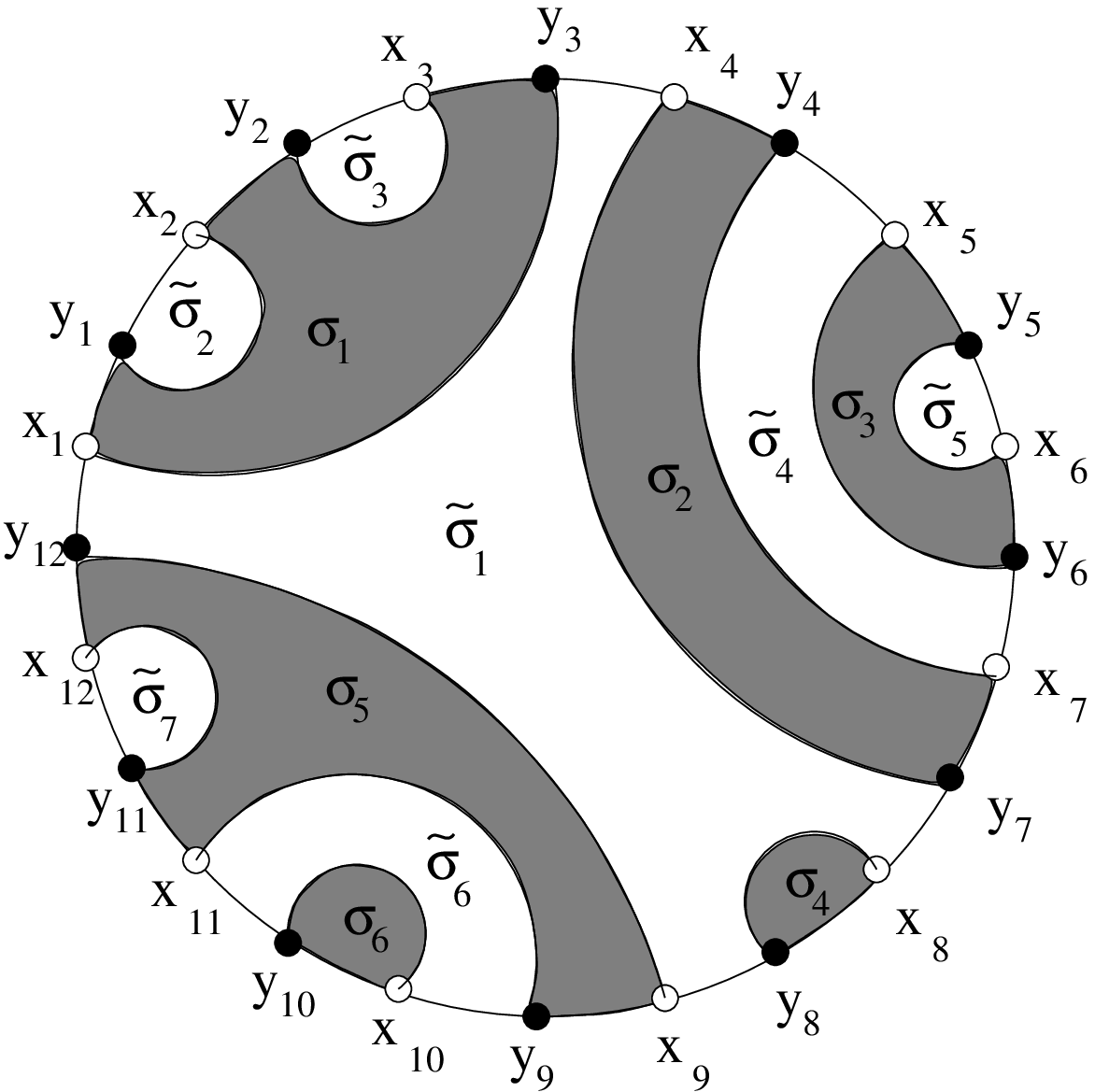}}
\end{array}
\eeq
where dark faces (resp. white faces) correspond to the cycles of $\sigma$ (resp. $S \sigma$).
For that permutation $C^{(12)}_\sigma$ is worth:
{\small
\beq
\begin{array}{l}
 F^{(3)}(x_1,y_3,x_3,y_2,x_2,y_1) F^{(2)}(x_5,y_6,x_6,y_5) F^{(3)}(x_9,y_{12},x_{12},y_{11},x_{11},y_9) \cr
\times F^{(2)}(x_4,y_7,x_7,y_4) F^{(4)}(x_1,y_3,x_4,y_7,x_8,y_8,x_9,y_{12}) F^{(2)}(x_5,y_6,x_7,y_4)\cr
\times F^{(2)}(x_{10},y_{10},x_{11},y_9)\cr
\end{array}
\eeq}

\subsubsection*{Example $k\leq 3$}

\beq
C^{(1)} = 1
\eeq
\beq
C^{(2)}_{Id} = {1\over (x_2-x_1)(y_2-y_1)}
\eeq
\beq
C^{(2)}_{(12)} = {1\over (x_2-x_1)(y_1-y_2)}
\eeq

\beq
C^{(3)}_{Id} = {1\over (x_1-x_3)}\left({1\over y_3-y_1}{1\over (x_2-x_3)(y_3-y_2)}+{1\over y_3-y_2}{1\over (x_1-x_2)(y_2-y_1)}\right)
\eeq
\beq
C^{(3)}_{(1)(23)} = {1\over (x_1-x_2)}{1\over y_3-y_1} {1\over (x_2-x_3)(y_2-y_3)}
\eeq
\beq
C^{(3)}_{(12)(3)} = {1\over (x_1-x_3)}{1\over y_3-y_2} {1\over (x_1-x_2)(y_1-y_2)}
\eeq
\beq
C^{(3)}_{(123)} = {1\over (x_1-x_2)(x_2-x_3)(y_1-y_2)}{1\over y_3-y_1} + {1\over (x_1-x_3)(x_1-x_2)(y_2-y_1)}{1\over y_3-y_2}
\eeq
\beq
C^{(3)}_{(13)(2)} = -{1\over y_3-y_1}{1\over (x_1-x_3)(x_2-x_3)(y_1-y_2)}
\eeq
\beq
C^{(3)}_{(132)} = 0
\eeq

\subsubsection*{Example: Rainbows}

The rainbow is the permutation
\beq
\sigma(j)=k+1-j
\eeq
if $k$ is even:
\bea
C^{(k)}_\sigma
&=& {1\over \prod_{i=1}^{k/2} (x_{k+1-i}-x_i)(y_i-y_{k+1-i})\prod_{i=1}^{k/2-1} (x_{k+1-i}-x_{i+1})(y_i-y_{k-i})} \cr
\eea
if $k$ is odd:
\bea
C^{(k)}_\sigma
&=& {1\over \prod_{i=1}^{(k-1)/2} (x_{k+1-i}-x_i)(x_{k+1-i}-x_{i+1})(y_i-y_{k+1-i})(y_i-y_{k-i})} \cr
\eea

\subsection{Properties of $C_\sigma$}

\bl\label{Ccyclic}
 The $C^{(k)}_\sigma$'s are cyclically invariant:
\beq
C^{(k)}_\sigma(x_1,y_1,x_2,y_2,\dots, x_k,y_k) = C^{(k)}_\sigma(x_2,y_2,\dots, x_k,y_k,x_1,y_1)
\eeq
\el
\proof{It follows from Lemma \ref{lemFcyclic}.}

\bl\label{lemCressimple}
 The $C^{(k)}_\sigma$'s have at most simple poles in all their variables and are such that:
\beq\label{CRessimple}
C^{(k)}_\sigma(x_1,y_1,x_2,y_2,\dots, x_k,y_k) = \sum_{j=1}^{k-1} {1\over y_k-y_j}
\Res_{y'_k\to y_j} C^{(k)}_\sigma(x_1,y_1,x_2,y_2,\dots, x_k,y'_k) dy'_k
\eeq
\el
\proof{If $\sigma$ is planar, the pair $(y_k,y_j)$  can appear in at most one factor of \eq{defCprodF},
and the results follows from Lemma \ref{lemFsimplepoles}.
}

\bt\label{ThrecC}
The $C^{(k)}_\sigma$'s, with $\sigma(1)\neq k$, satisfy the recursion formula \eq{Crec}:
\bea\label{Crecth}
C^{(k)}_\sigma &=&
{1\over x_{\sigma^{-1}(k)}-x_1}
\sum_{j=1}^{k-1}
\sum_{\tau\in \Sym(1,\dots,j)}
\sum_{\rho\in \Sym(j+1,\dots,k)}\,
\delta_{\sigma,\tau\rho}\,\,
{1\over y_k-y_j} C^{(j)}_\tau C^{(k-j)}_\rho
\eea
\et

\proof{

Let $\pi$ be the cycle of length $l+1\geq 2$ of $S\sigma$ which contains $x_1$ and $y_k$:
\beq
\pi = (x_1 \to y_{i_1-1} \to x_{i_1} \to y_{i_2-1}\to \dots \to y_{i_{l-1}} \to x_{i_l} \to y_k \to x_1)
\eeq
\beq
i_{j+1}:=\sigma(i_j)+1 \virg i_0:=1
\eeq
Planarity implies that:
\beq
i_0<i_1<i_2<\dots <i_l
\eeq
Since $\sigma$ is planar, there exists a unique way of factorizing $\sigma$ as:
\beq
\sigma= \prod_{j=0}^l \sigma_j
\virg
\sigma_j\in \Symp(i_{j},\dots,i_{j+1}-1)
\eeq

From the definition of $C^{(k)}_\sigma$ we have:
\bea\label{DefC}
C^{(k)}_\sigma = F^{(l+1)}(x_1,y_{i_1-1},x_{i_1},y_{i_2-1},\dots,y_{i_l-1},x_{i_l},y_k) \prod_{j} C^{(i_{j+1}-i_j)}_{\sigma_j}(x_{i_{j}},\dots, y_{i_{j+1}-1}) \cr
\eea
and using \eq{recF}, we have:
\beq
\begin{array}{l}
C^{(k)}_\sigma
= F^{(l+1)}(x_1,y_{i_1-1},x_{i_1},y_{i_2-1},\dots,y_{i_l-1},x_{i_l},y_k) \prod_{j} C^{(i_{j+1}-i_j)}_{\sigma_j}(x_{i_{j}},\dots, y_{i_{j+1}-1}) \cr
= \sum_{m=1}^l {F^{(m)}(x_1,y_{i_1-1},x_{i_1},\dots,x_{i_{m-1}},y_{i_m-1}) F^{(l+1-m)}(x_{i_m},y_{i_{m+1}-1},x_{i_{m+1}},\dots,x_{i_l},y_k)\over (x_{i_l}-x_1)(y_k-y_{i_m-1})} \cr
\qquad \prod_{j} C^{(i_{j+1}-i_j)}_{\sigma_j}(x_{i_{j}},\dots, y_{i_{j+1}-1})  \cr
\end{array}
\eeq
notice that $i_{l}=\sigma^{-1}(k)$, and note:
\beq
\tau_m := \prod_{j=1}^{m-1} \sigma_j
\virg
\rho_m := \prod_{j=m}^{l} \sigma_j
\eeq
We have:
\beq
\tau_m\in\Symp(1,\dots,i_m-1)
\virg
\rho_m\in\Symp(i_m,\dots,k)
\eeq
Eq. (\ref{DefC}) gives:
\bea
C^{(k)}_\sigma &=&
{1\over x_{\sigma^{-1}(k)}-x_1}
\sum_{m=1}^{l}
{1\over y_k-y_{i_m-1}} C^{(i_m)}_{\tau_m} C^{(k-i_m)}_{\rho_m}
\eea
It is clear, from the planarity condition that if there exists some $j$ and $\tau$ and $\rho$, such that:
\beq
\sigma=\tau\rho
\virg \tau\in \Symp(1,\dots,j)
\virg \rho\in \Symp(j+1,\dots,k)
\eeq
then, one must have $j=i_m$, $\tau=\tau_m$ and $\rho=\rho_m$ for some $m$.
}

\bl\label{lemcancelpoleC}
For any transposition $(k,j)$ (with $k\neq j$), we have:
\beq\label{cancelpoleC}
\Res_{y'_k\to y_j} C^{(k)}_\sigma(x_1,y_1,x_2,y_2,\dots, x_k,y'_k) dy'_k
=
- \Res_{y'_k\to y_j} C^{(k)}_{(k,j)\circ\sigma}(x_1,y_1,x_2,y_2,\dots, x_k,y'_k) dy'_k
\eeq
\el
\proof{
It is trivial if $y_j$ and $y_k$ are not in the same face, because both sides vanish: the LHS has no pole and the RHS is a non-planar permutation.
The case where $y_j$ and $y_k$ belong to the same face reduces to proving the Lemma for $\sigma=Id$.

For $\sigma=Id$, we prove it by recursion on $k$. It clearly works for $k=1$ and $k=2$.
Assume that it works up to $k-1$.

From the definition \eq{defC} we have:
\bea
&& C^{(k)}_{(k,j)}(x_1,y_1,x_2,y_2,\dots, x_k,y_k) \cr
&=& F^{(j)}(x_1,y_1,\dots,x_j,y_k)F^{(k-j)}(x_{j+1},y_{j+1},\dots,y_{k-1},x_k,y_j)F^{(2)}(x_j,y_k,x_k,y_j) \cr
&=& {F^{(j)}(x_1,y_1,\dots,x_j,y_k)F^{(k-j)}(x_{j+1},y_{j+1},\dots,y_{k-1},x_k,y_j)\over (x_j-x_k)(y_k-y_j)} \cr
\eea
and thus:
\beq\label{eqCtauRes}
\Res_{y'_k\to y_j} C^{(k)}_{(k,j)}(x_1,y_1,\dots, x_k,y_k)
= {F^{(j)}(x_1,y_1,\dots,x_j,y_j)F^{(k-j)}(x_{j+1},y_{j+1},\dots,x_k,y_j)\over (x_j-x_k)}
\eeq

On the LHS, we have from \eq{recF}:
\bea
&& \Res_{y'_k\to y_j} F^{(k)}(x_1,y_1,x_2,y_2,\dots, x_k,y'_k) \cr
&=& \Res_{y'_k\to y_j}   \sum_{l=1}^{k-1} {F^{(l)}(x_1,y_1,x_2,\dots,x_l,y_l) F^{(k-l)}(x_{l+1},y_{l+1},\dots,x_k,y'_k) \over (x_k-x_1)(y'_k-y_l)} \cr
&=&   {F^{(j)}(x_1,y_1,x_2,\dots,x_j,y_j) F^{(k-j)}(x_{j+1},y_{j+1},\dots,x_k,y_j) \over (x_k-x_1)} \cr
&& +  \sum_{l=1}^{j-1} {F^{(l)}(x_1,y_1,x_2,\dots,x_l,y_l)  \over (x_k-x_1)(y_j-y_l)} \Res_{y'_k\to y_j}   F^{(k-l)}(x_{l+1},y_{l+1},\dots,x_k,y'_k) \cr
\eea
Then, from the recursion hypothesis, and from \eq{eqCtauRes} we have:
\bea
&& \Res_{y'_k\to y_j} F^{(k)}(x_1,y_1,x_2,y_2,\dots, x_k,y'_k) \cr
&=&   {F^{(j)}(x_1,y_1,x_2,\dots,x_j,y_j) F^{(k-j)}(x_{j+1},y_{j+1},\dots,x_k,y_j) \over (x_k-x_1)} \cr
&& -  \sum_{l=1}^{j-1} {F^{(l)}(x_1,y_1,x_2,\dots,x_l,y_l)  \over (x_k-x_1)(y_j-y_l)} \Res_{y'_k\to y_j}   C_{(k,j)}^{(k-l)}(x_{l+1},y_{l+1},\dots,x_k,y'_k) \cr
&=&   {F^{(j)}(x_1,y_1,x_2,\dots,x_j,y_j) F^{(k-j)}(x_{j+1},y_{j+1},\dots,x_k,y_j) \over (x_k-x_1)} \cr
&& -  \sum_{l=1}^{j-1} {F^{(l)}(x_1,y_1,x_2,\dots,x_l,y_l)  \over (x_k-x_1)(y_j-y_l)} \cr
&& \qquad {F^{(j-l)}(x_{l+1},y_{l+1},\dots,x_j,y_j)F^{(k-j)}(x_{j+1},y_{j+1},\dots,y_{k-1},x_k,y_j)\over (x_j-x_k)} \cr
\eea
In the last line, we use again \eq{recF} and get:
\bea\label{eqCIdRes2}
&& \Res_{y'_k\to y_j} F^{(k)}(x_1,y_1,x_2,y_2,\dots, x_k,y'_k) \cr
&=&   {F^{(j)}(x_1,y_1,x_2,\dots,x_j,y_j) F^{(k-j)}(x_{j+1},y_{j+1},\dots,x_k,y_j) \over (x_k-x_1)} \cr
&& -  (x_j-x_1) {F^{(j)}(x_1,y_1,x_2,\dots,x_j,y_j)  \over (x_k-x_1)} {F^{(k-j)}(x_{j+1},y_{j+1},\dots,y_{k-1},x_k,y_j)\over (x_j-x_k)} \cr
\eea
One can then see that the sum of \eq{eqCtauRes} and \eq{eqCIdRes2} vanishes, and the recursion hypothesis is proven for $k$.
}

\bl
\beq
\forall\, k>1\, , \qquad
\sum_{\sigma\in \Symp_k} C^{(k)}_\sigma(x_1,y_1,x_2,y_2,\dots, x_k,y_k) =0
\eeq
\el
\proof{
This expression is a rational function of all its variables.
Consider the poles at $y_k=y_j$, and write $\tau=(k,j)$.
One can split the symmetric group $\Symp_k$ into its two conjugacy classes wrt the subgroup generated by $\tau$: $\Symp_k=[Id]\oplus[\tau]$.
In other words:
\bea
&& \sum_{\sigma\in \Symp_k} C_\sigma(x_1,y_1,x_2,y_2,\dots, x_k,y_k) \cr
&=& \sum_{\sigma\in \Symp_k/\tau} C_\sigma(x_1,y_1,x_2,y_2,\dots, x_k,y_k)+C_{\tau\sigma}(x_1,y_1,x_2,y_2,\dots, x_k,y_k) \cr
\eea
From Lemma \ref{lemcancelpoleC}, the terms in the RHS have no pole at $y_k=y_j$.
Similarly, using cyclicity and doing the same for the $x$'s, we prove the lemma.
}

\medskip

The Lemmas we have just proven, are sufficient to prove the main theorem \ref{mainth}.
This is done in section \ref{proofcorrel}.

\subsection{Computation of the rational functions $F^{(k)}(x_1,y_1,\dots,x_k,y_k)$}

Although the exact computation of the $F^{(k)}$'s is not necessary for proving theorem \ref{mainth},
we do it for completeness.
In this section we give an explicit (and non-recursive) formula for the $F^{(k)}$'s.

A practical way of computing these formulas is described in Appendix A.

\bd\label{deff}
To every permutation $\sigma\in\Symp_{k-1}$, we associate a weight $f_\sigma$ computed as follows:
 \beq f_\sigma = \prod_{n=1}^l
\prod_{j=2}^{l_n}g_{i_{n,1},i_{n,j},i_{n,j+1}}\prod_{n=2}^{\td{l}}
\prod_{j=2}^{\td{l_n}}g_{\td{i}_{n,j},\td{i}_{n,1},\sigma (\td{i}_{n,j})}
\prod_{j=1}^{\td{l_1}}g_{\td{i}_{1,j},k,\sigma (\td{i}_{1,j})} \eeq
where
$g_{i,h,j}$ is defined as follows:
\beq
g_{i,h,j}:=\frac{1}{x_h-x_i} \,\frac{1}{y_h-y_j}
\eeq
and $\sigma$ and $S\sigma$ are decomposed into their product of cycles as in Definition \ref{defC}.
\ed

\bt
$F^{(k)}(x_1,y_1, \dots , x_k,y_k)$ is obtained as the sum of the
weights $f_\sigma$'s over all $\sigma \in \Sym_{k-1}$:
\beq
F^{(k)}(x_1,y_1, \dots , x_k,y_k) = \sum_{\sigma \in \Sym_{k-1} } f_\sigma
\eeq
\et

\proof{
First of all, let us interpret diagrammatically the recursion relation \eq{recF} defining the F's:
\beq
\begin{array}{r}
{\epsfxsize 14cm\epsffile{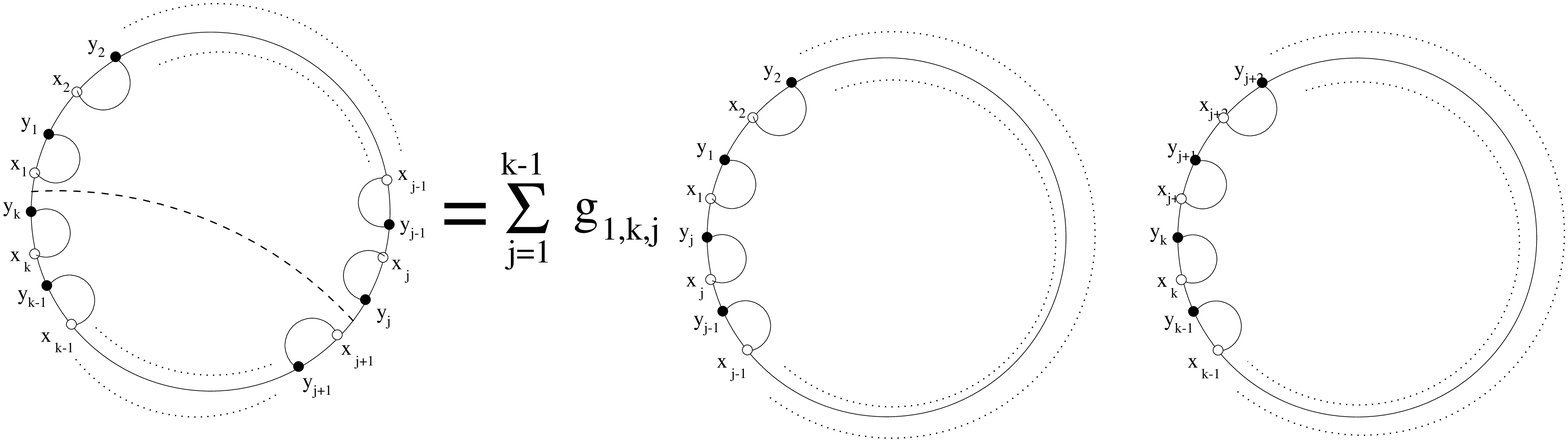}}
\end{array}
\eeq

Actually, this recursion relation is nothing else but a rule for cutting a graph along the dashed line into two smaller ones.
The weight of a graph is then obtained as the sum over all the possible ways of cutting it in two.

Notice that $F^{(k)}$ is the sum of $\Cat(k-1)$ different terms.

Let us now explicit this bijection with the graphs with $k-1$ arches. In order to compute one of the terms composing
$F^{(k)}$, one has to cut it with the help of the recursion relation until one obtains only graphs with one arch.
That is to say that one cuts it $k-1$ times along non intersecting lines (corresponding to the dashed one
in the recursion relation). If one draws these cutting lines on the circle, one obtains a graph with $k-1$ arches dual of the original one.
Thus every way of cutting a graph with $k-1$ arches is associated to a planar permutation $\sigma \in \overline{\Sigma} (i \dots k-1)$. Let us now prove
that the term obtained by this cutting is equal to $f_\sigma$.

For the sake of simplicity, one denotes the identity graph of $(x_j,y_j, \dots , x_k,y_k)$
by circle $(j,j+1, \dots, k)$. In these conditions, the recursion relation reads:
\beq
(1,2, \dots, k) = \sum_{j=1}^{k-1} g_{1,k,j} (1, \dots , j) (j+1,\dots ,k)
\eeq

Let $\sigma$ be a permutation of $(1, \dots , k-1)$. Cut it along the line going from the boundary $(x_1,y_{k})$ to
$(y_{\sigma(1)},x_{S \sigma(1)})$. It results from this operation the factor $g_{1,k,\sigma(1)}$ and the circles
$(1, \dots , \sigma(1))$ and $(S \sigma(1), \dots, k)$:
\beq
(1,\dots ,k) \rightarrow^{\sigma} g_{1,k,\sigma(1)} (1, \dots , \sigma(1)) (S \sigma(1), \dots, k)
\eeq

Then cutting the circle $(S \sigma(1), \dots, k)$ along $(y_k,x_{S \sigma(1)}) \rightarrow (y_{\sigma S \sigma(1)}, x_{S \sigma S \sigma(1)})$
gives:
\beq
(S \sigma(1), \dots, k) \rightarrow^{\sigma} g_{S \sigma(1),k, \sigma S \sigma(1)} (S \sigma(1), \dots , \sigma S \sigma(1)) (S \sigma S \sigma(1), \dots , k)
\eeq

One pursues this procedure step by step by always cutting the circle containing k. Using the former notations, this reads:
\beq
(1,\dots ,k) \rightarrow^{\sigma} \prod_{j=1}^{\td{l}_1} g_{\td{i}_{1,j}, k, \sigma (\td{i}_{1,j})}
(\td{i}_{1,j}, \dots , \sigma(\td{i}_{1,j}))
\eeq

So one has computed the weight associated to the first $S\sigma$ - cycle. The remaining circles correspond to
$\sigma$-cycles. Let us compute their weight by considering for example $(\td{i}_{1,1},\dots , \sigma(\td{i}_{1,1})) = (i_{1,1}, \dots , i_{1,2})$.

The cut along the line $(x_{i_{1,1}},y_{i_{1,2}}) \rightarrow (y_{i_{1,3}},x_{S(i_{1,1})})$ gives:
\beq
(i_{1,1}, \dots , i_{1,2}) \rightarrow^{\sigma} g_{i_{1,1},i_{1,2},i_{1,3}} (i_{1,1}, \dots , i_{1,3}) (S(i_{1,3}), \dots , i_{1,2})
\eeq

Keeping on cutting the circle containing $i_{1,1}$ at every step gives:
\beq
(i_{1,1}, \dots , i_{1,2}) \rightarrow^{\sigma} \prod_{j=2}^{l_1} g_{i_{1,1},i_{1,j},i_{1,j+1}} (S(i_{1,j+1}), \dots , i_{1,j})
\eeq

One can notice that the remaining circles in the RHS correspond to cycles of $S\sigma$ whose contribution has not been taken
into account yet. One can then compute their values by
following a procedure similar to the one used for the first $S\sigma$-cycle.

One can then recursively cut the circles so that one finally obtains only circles containing only one element.
This recursion is performed by alternatively processing on $\sigma$-cycles and $S\sigma$-cycles.

Thus, one straightforwardly finds:
\beq
(1,\dots ,k) \rightarrow^{\sigma} \prod_{n=1}^l
\prod_{j=2}^{l_n}g_{i_{n,1},i_{n,j},i_{n,j+1}}\prod_{n=2}^{\td{l}}
\prod_{j=2}^{\td{l_n}}g_{\td{i}_{n,j},\td{i}_{n,1},\sigma (\td{i}_{n,j})}
\prod_{j=1}^{\td{l_1}}g_{\td{i}_{1,j},k,\sigma (\td{i}_{1,j})} = f_\sigma
\eeq

And then:
\beq
F^{(k)} = \sum_{\sigma\in \Symp_{k-1}} f_\sigma
\eeq
}

{\bf Example:}
Let us compute the weight associated to the permutation $\sigma \in \Symp_{12}$ introduced earlier. Starting from the circle $(1, \dots , 13)$, one will proceed
step by step the following cutting:
\beq
\begin{array}{r}
{\epsfxsize 6cm\epsffile{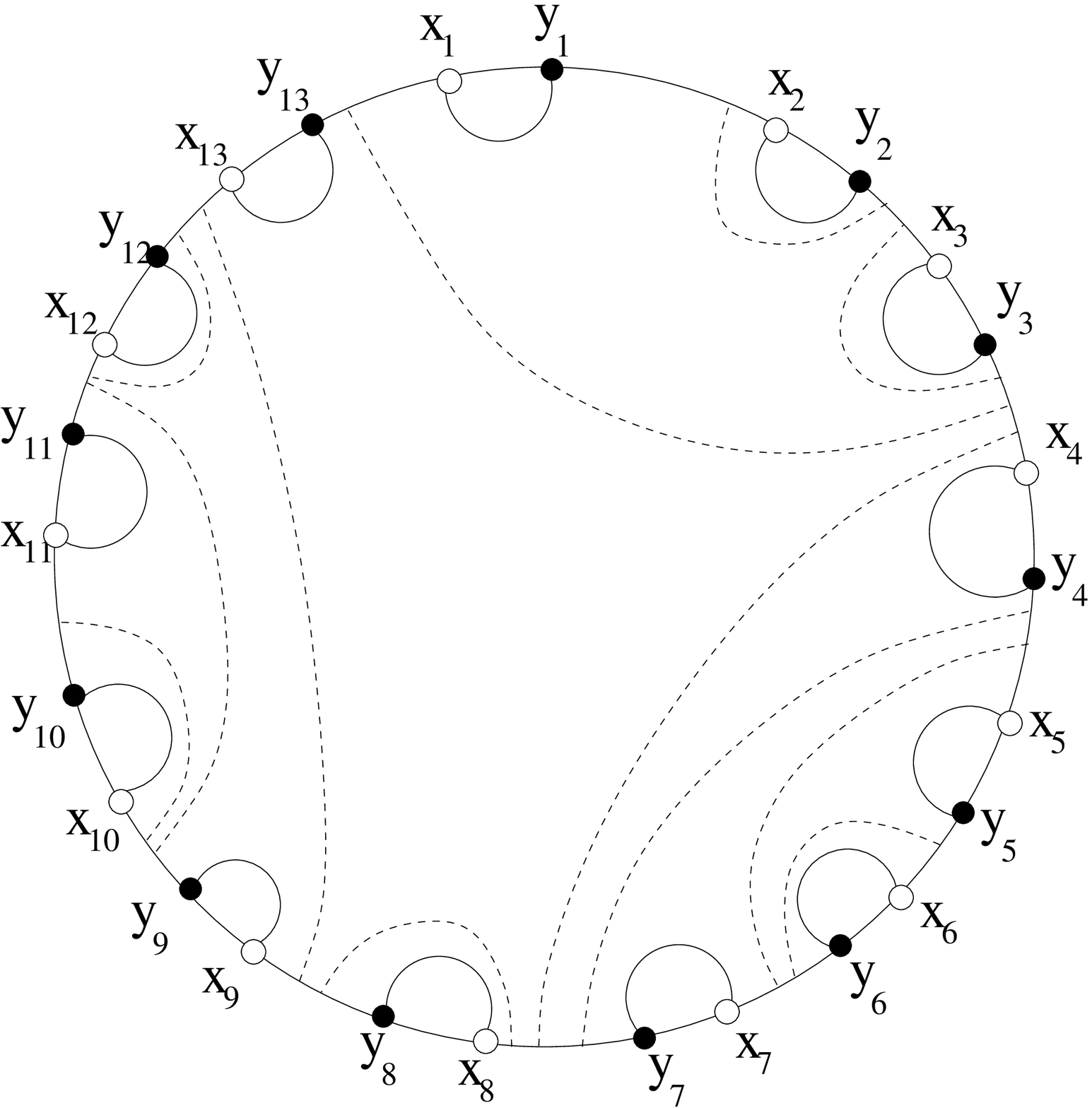}}
\end{array}
\eeq

The first step consists in cutting along the $\td{\sigma}_1$ cycle. The dashed lines show where one cuts the circles.
 Note that one do not represent the circles of unite length. The associated weight is $g_{1,13,3} \,\, g_{4,13,7} \,\, g_{8,13,8} \,\, g_{9,13,12}$.

The second step consists in cutting along the remaining $\sigma$ cycles. One associates the weight $g_{1,3,2} \,\, g_{1,2,1} \times g_{4,7,4} \times g_{9,12,11} \,\, g_{9,11,9}$ to this step.

The weights associated to the two last cuttings are $g_{5,7,6} \times g_{10,11,10}$ and $g_{5,6,5}$.

\beq
\begin{array}{r}
{\epsfxsize 13.4cm\epsffile{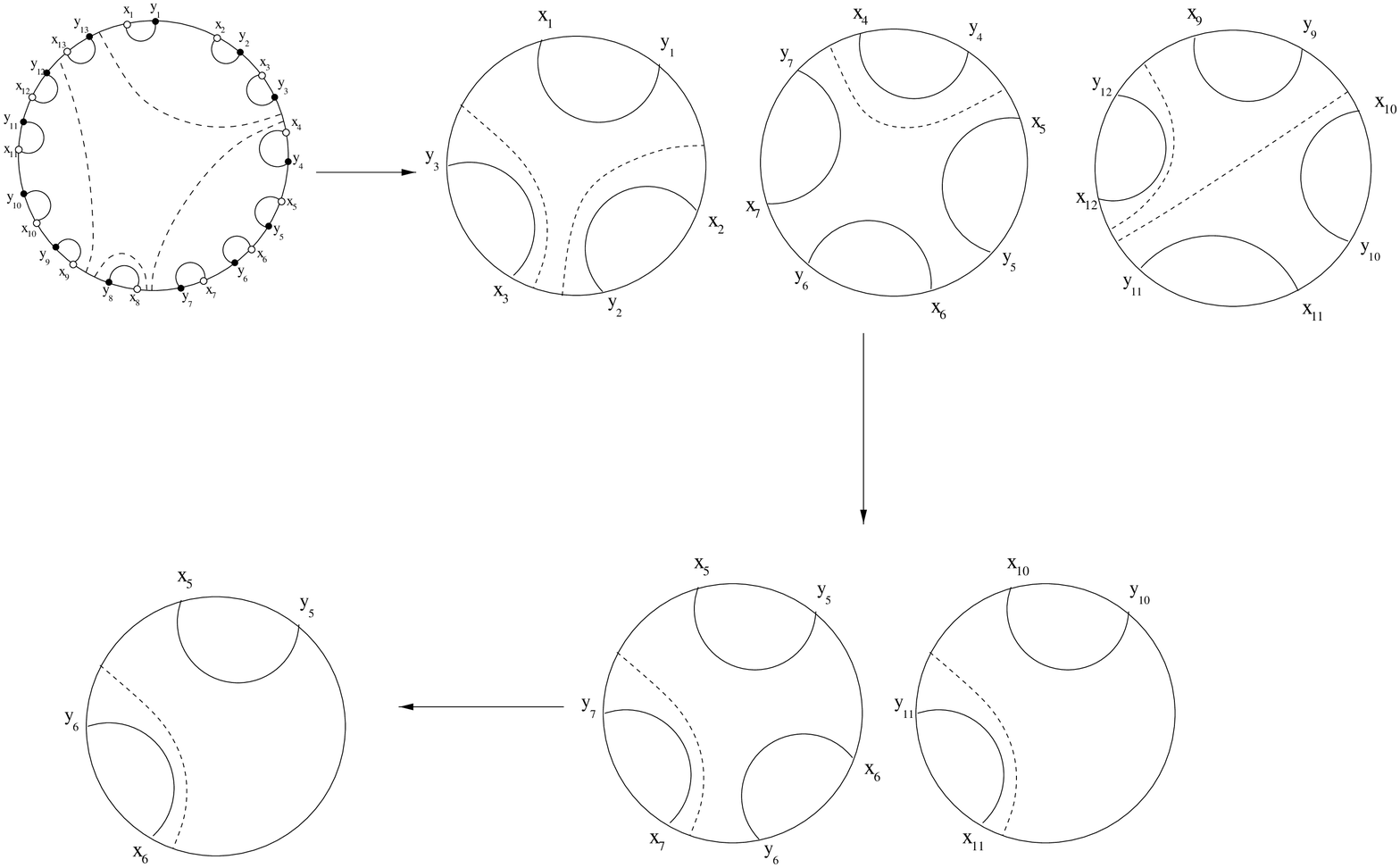}}
\end{array}
\eeq

Finally, the weight of this planar permutation is then:
\beq
f_{\sigma} = g_{1,13,3} \,\, g_{4,13,7} \,\, g_{8,13,8} \,\, g_{9,13,12} \,\, g_{1,3,2} \,\, g_{1,2,1} \,\,  g_{4,7,4} \,\, g_{9,12,11} \,\, g_{9,11,9} \,\, g_{5,7,6} \,\, g_{10,11,10} \,\, g_{5,6,5}
\eeq

\subsection{Proof of the main theorem}
\label{proofcorrel}
We now prove that the function $\widehat{W}$ defined by the RHS of \eq{Ansatz} and the functions $\widehat{U}$ and $\widehat{P}$ defined in
theorem \ref{thansatzloop} satisfy the system of equations \eq{loopeqU} and \eq{loopeqW}.

{\noindent \bf Proof of theorem \ref{thansatzloop}:}

Using \eq{CRessimple}, one has:
\bea
&&\widehat{U}_k(x_1,\dots,y_k)
= \Pol_{y_k} V'_2(y_k)\,\widehat{W}_k(x_1,y_1,\dots,x_k,y_k) \cr
&&=  \Pol_{y_k} V'_2(y_k)\,\sum_{\sigma\in \Sym_k}\, C^{(k)}_\sigma(x_1,y_1,\dots,x_k,y_k)\,\, \prod_{i=1}^k W_1(x_i,y_{\sigma(i)}) \cr
&&=   \sum_{\sigma\in \Sym_k}\,\sum_{j\neq k}\,\Res_{y'_k\to y_j}  C^{(k)}_\sigma(x_1,y_1,\dots,x_k,y'_k) dy'_k \,\, \prod_{i=1}^{k-1} W_1(x_{\sigma^{-1}(i)},y_i) \cr
&& \quad \Pol_{y_k}{V'_2(y_k)W_1(x_{\sigma^{-1}(k)},y_k)\over y_k-y_j} \cr
&&=   \sum_{\sigma\in \Sym_k}\,\sum_{j\neq k}\,\Res_{y'_k\to y_j}  C^{(k)}_\sigma(x_1,y_1,\dots,x_k,y'_k) dy'_k \,\, \prod_{i=1}^{k-1} W_1(x_{\sigma^{-1}(i)},y_i) \cr
&& \quad \Pol_{y_k}{(\td{W}(y_k)-X(y_k))W_1(x_{\sigma^{-1}(k)},y_k)\over y_k-y_j} \cr
&&=  - \sum_{\sigma\in \Sym_k}\,\sum_{j\neq k}\,\Res_{y'_k\to y_j}  C^{(k)}_\sigma(x_1,y_1,\dots,x_k,y'_k) dy'_k \,\, \prod_{i=1}^{k-1} W_1(x_{\sigma^{-1}(i)},y_i) \cr
&& \quad \Pol_{y_k}{X(y_k)W_1(x_{\sigma^{-1}(k)},y_k)\over y_k-y_j} \cr
&&=  \sum_{\sigma\in \Sym_k}\,\sum_{j\neq k}\,\Res_{y'_k\to y_j}  C^{(k)}_\sigma(x_1,y_1,\dots,x_k,y'_k) dy'_k \,\, \prod_{i=1}^{k-1} W_1(x_{\sigma^{-1}(i)},y_i) \cr
&& \Pol_{y_k} {(x_{\sigma^{-1}(k)}-X(y_k))W_1(x_{\sigma^{-1}(k)},y_k)-(x_{\sigma^{-1}(k)}-X(y_j))W_1(x_{\sigma^{-1}(k)},y_j)\over y_k-y_j} \cr
&&=  \sum_{\sigma\in \Sym_k}\,\sum_{j\neq k}\,\Res_{y'_k\to y_j}  C^{(k)}_\sigma(x_1,y_1,\dots,x_k,y'_k) dy'_k \,\, \prod_{i=1}^{k-1} W_1(x_{\sigma^{-1}(i)},y_i) \cr
&&  {(x_{\sigma^{-1}(k)}-X(y_k))W_1(x_{\sigma^{-1}(k)},y_k)-(x_{\sigma^{-1}(k)}-X(y_j))W_1(x_{\sigma^{-1}(k)},y_j)\over y_k-y_j} \cr
\eea

Indeed, using \eq{W1U1}, one sees that the last expression is a polynomial in $y_k$:
\bea
&& {(x_{\sigma^{-1}(k)}-X(y_k))W_1(x_{\sigma^{-1}(k)},y_k)-(x_{\sigma^{-1}(k)}-X(y_j))W_1(x_{\sigma^{-1}(k)},y_j)\over y_k-y_j} \cr
&=& {{E(x_{\sigma^{-1}(k)},y_k)\over y_k-Y(x_{\sigma^{-1}(k)})}-{E(x_{\sigma^{-1}(k)},y_j)\over y_j-Y(x_{\sigma^{-1}(k)})}\over y_k-y_j} \cr
\eea

We have to check \eq{loopeqW}, i.e. that $A=0$ with:
\bea\label{defA}
A&:=& \sum_{j} {\widehat{W}_{k-j}(x_{j+1},\dots,y_k)-\widehat{W}_{k-j}(x_{j+1},\dots,x_k,y_j)\over y_k-y_j}\, \widehat{W}_j(x_1,\dots,y_j) \cr
&& - \widehat{U}_k(x_1,\dots,y_k)   \cr
&& + (x_1-X(y_k))\,\widehat{W}_k(x_1,y_1,x_2,\dots,x_k,y_k) \cr
\eea
We have:
\bea
A&=& \sum_{j\neq k} \sum_\pi\sum_\tau { C^{(j)}_\tau(x_1,\dots,y_j)\,C^{(k-j)}_\pi(x_{j+1},\dots,y_k) \over y_k-y_j} \cr
&& \quad \times \prod_{i=1}^{j} W_1(x_{\tau^{-1}(i)},y_i)\,\prod_{i=j+1}^{k} W_1(x_{\pi^{-1}(i)},y_i) \cr
&& - \sum_{j\neq k} \sum_\pi\sum_\tau { C^{(j)}_\tau(x_1,\dots,y_j)\,C^{(k-j)}_\pi(x_{j+1},\dots,y_j) \over y_k-y_j}\,  W_1(x_{\pi^{-1}(k)},y_j) \,\cr
&& \quad \times \prod_{i=1}^{j} W_1(x_{\tau^{-1}(i)},y_i)\,\prod_{i=j+1}^{k-1} W_1(x_{\pi^{-1}(i)},y_i) \cr
&& - \sum_\sigma \sum_{j\neq k} {(x_{\sigma^{-1}(k)}-X(y_k))\over y_k-y_j}\,W_1(x_{\sigma^{-1}(k)},y_k) \cr
&& \quad \times \left(\Res_{y_k\to y_j} C^{(k)}_\sigma\right) \prod_{i=1}^{k-1} W_1(x_{\sigma^{-1}(i)},y_i) \cr
&& + \sum_\sigma \sum_{j\neq k} {(x_{\sigma^{-1}(k)}-X(y_j))\over y_k-y_j} \,W_1(x_{\sigma^{-1}(k)},y_j) \cr
&& \quad \times \left(\Res_{y_k\to y_j} C^{(k)}_\sigma\right) \prod_{i=1}^{k-1} W_1(x_{\sigma^{-1}(i)},y_i) \cr
&& + (x_1-X(y_k)) \, \sum_\sigma C^{(k)}_\sigma(x_1,\dots,y_k) W_1(x_{\sigma^{-1}(k)},y_k) \prod_{i=1}^{k-1} W_1(x_{\sigma^{-1}(i)},y_i)  \cr
\eea
Using \eq{CRessimple} in the last line, adding it to the 4th line, and using \eq{CRessimple} again, we get:
\bea
A&=& \sum_{j\neq k} \sum_\pi\sum_\tau { C^{(j)}_\tau(x_1,\dots,y_j)\,C^{(k-j)}_\pi(x_{j+1},\dots,y_k) \over y_k-y_j} \cr
&& \quad \times \prod_{i=1}^{j} W_1(x_{\tau^{-1}(i)},y_i)\,\prod_{i=j+1}^{k} W_1(x_{\pi^{-1}(i)},y_i) \cr
&& + (x_1-x_{\sigma^{-1}(k)}) \, \sum_\sigma C^{(k)}_\sigma(x_1,\dots,y_k) W_1(x_{\sigma^{-1}(k)},y_k) \prod_{i=1}^{k-1} W_1(x_{\sigma^{-1}(i)},y_i)  \cr
&& - \sum_{j\neq k} \sum_\pi\sum_\tau { C^{(j)}_\tau(x_1,\dots,y_j)\,C^{(k-j)}_\pi(x_{j+1},\dots,y_j) \over y_k-y_j}\,  W_1(x_{\pi^{-1}(k)},y_j) \,\cr
&& \qquad \prod_{i=1}^{j} W_1(x_{\tau^{-1}(i)},y_i)\,\prod_{i=j+1}^{k-1} W_1(x_{\pi^{-1}(i)},y_i) \cr
&& + \sum_\sigma \sum_{j\neq k} {(x_{\sigma^{-1}(k)}-X(y_j))\over y_k-y_j} \,W_1(x_{\sigma^{-1}(k)},y_j) \cr
&& \quad \times \left(\Res_{y_k\to y_j} C^{(k)}_\sigma\right) \prod_{i=1}^{k-1} W_1(x_{\sigma^{-1}(i)},y_i) \cr
\eea
Using \eq{Crec} in the second line, exactly cancels the first line, and thus we get:
\bea
A&=& - \sum_{j\neq k} \sum_\pi\sum_\tau { C^{(j)}_\tau(x_1,\dots,y_j)\,C^{(k-j)}_\pi(x_{j+1},\dots,y_j) \over y_k-y_j}\,  W_1(x_{\pi^{-1}(k)},y_j) \,\cr
&& \qquad \prod_{i=1}^{j} W_1(x_{\tau^{-1}(i)},y_i)\,\prod_{i=j+1}^{k-1} W_1(x_{\pi^{-1}(i)},y_i) \cr
&& + \sum_\sigma \sum_{j\neq k} {(x_{\sigma^{-1}(k)}-X(y_j))\over y_k-y_j} \,W_1(x_{\sigma^{-1}(k)},y_j) \cr
&& \quad \times \left(\Res_{y_k\to y_j} C^{(k)}_\sigma\right) \prod_{i=1}^{k-1} W_1(x_{\sigma^{-1}(i)},y_i) \cr
\eea
which is a rational fraction in $y_k$ with poles at $y_k=y_j$ for some $j$.
From Lemma \ref{lemcancelpoleC}, $A$ as defined in \eq{defA} cannot have poles at $y_k=y_j$, thus $A=0$.

\bigskip
Now, we have to check \eq{loopeqU}
Using \eq{CRessimple}, one has:
\bea
&& \widehat{P}_k(x_1,\dots,y_k) - \widehat{W}_{k-1}(x_2,\dots,x_{k},y_1) \cr
&=& \Pol_{x_1} V'_1(x_1)\,\widehat{U}_k(x_1,y_1,\dots,x_k,y_k) \cr
&=&  \Pol_{x_1} Y(x_1)  \sum_{\sigma\in \Sym_k}\,\sum_{j\neq k}\,\Res_{y'_k\to y_j}  C^{(k)}_\sigma \,\, \prod_{i=1}^{k-1} W_1(x_{\sigma^{-1}(i)},y_i) \cr
&& \quad  {U_1(x_{\sigma^{-1}(k)},y_k)-U_1(x_{\sigma^{-1}(k)},y_j)\over y_k-y_j} \cr
&=&  \Pol_{x_1} Y(x_1)  \sum_{\sigma\in \Sym_k,\,\sigma(1)=k}\,\sum_{j\neq k}\,\Res_{y'_k\to y_j}  C^{(k)}_\sigma \,\, \prod_{i=1}^{k-1} W_1(x_{\sigma^{-1}(i)},y_i) \cr
&& \quad  {U_1(x_1,y_k)-U_1(x_1,y_j)\over y_k-y_j} \cr
&& + \Pol_{x_1} Y(x_1)  \sum_{\sigma\in \Sym_k,\,\sigma(1)\neq k}\,\sum_{j\neq k}\,\Res_{y'_k\to y_j}  C^{(k)}_\sigma \,\, W_1(x_1,y_{\sigma(1)})\prod_{i\neq k,\sigma(1)} W_1(x_{\sigma^{-1}(i)},y_i) \cr
&& \quad  {U_1(x_{\sigma^{-1}(k)},y_k)-U_1(x_{\sigma^{-1}(k)},y_j)\over y_k-y_j} \cr
&=&  -  \sum_{\sigma\in \Sym_k,\,\sigma(1)=k}\,\sum_{j\neq k}\,\sum_{l\neq 1}\,\Res_{x_1\to x_l}\,\Res_{y'_k\to y_j}  C^{(k)}_\sigma \,\, \prod_{i=1}^{k-1} W_1(x_{\sigma^{-1}(i)},y_i) \cr
&& \quad  {E(x_1,y_k)-E(x_1,y_j)-E(x_l,y_k)+E(x_l,y_j)\over (x_1-x_l)(y_k-y_j)} \cr
&& +   \sum_{\sigma\in \Sym_k,\,\sigma(1)\neq k}\,\sum_{j\neq k}\,\sum_{l\neq 1}\,\Res_{x_1\to x_l}\,\Res_{y'_k\to y_j}  C^{(k)}_\sigma \,\,\prod_{i\neq k,\sigma(1)} W_1(x_{\sigma^{-1}(i)},y_i) \cr
&& \quad  {(y_{\sigma(1)}-Y(x_1))W_1(x_1,y_{\sigma(1)})-(y_{\sigma(1)}-Y(x_l))W_1(x_l,y_{\sigma(1)})\over x_1-x_l} \cr
&&  \quad  {U_1(x_{\sigma^{-1}(k)},y_k)-U_1(x_{\sigma^{-1}(k)},y_j)\over y_k-y_j} \cr
\eea

In order to satisfy \eq{loopeqU}, we must prove that $B=0$, where:
\bea
B&:=& \sum_{l=2}^k {\widehat{W}_{l-1}(x_1,y_1,\dots,y_{l-1})-\widehat{W}_{l-1}(x_l,y_1,x_2,\dots,y_{l-1})\over x_1-x_l}\cr
&& \quad \times \widehat{U}_{k-l+1}(x_l,y_l,\dots,x_k,y_k) \cr
&& +\sum_{l=1}^{k-1} \widehat{W}_l(x_1,\dots,y_l)\,\widehat{W}_{k-l}(x_k,y_l,\dots,y_{k-1}) \cr
&& - \widehat{P}_k(x_1,y_1,x_2,\dots,x_k,y_k) \cr
&& - (Y(x_1)-y_k)\, \widehat{U}_k(x_1,\dots,y_k) \cr
\eea

One does it in a way very similar to the previous one, i.e. first prove, using \eq{Crec}, that $B$ is a rational fraction of $x_1$, with poles at $x_1=x_l$.
But $B$ can have no pole at $x_1=x_l$, so,$B=0$.

{$\bullet$}

\newsection{Matrix form of correlation functions}

So far, we have computed mixed correlation functions with only one trace, i.e. the generating function of connected discrete surfaces with
one boundary.
In this section, we generalize this theory to the computation of generating functions of non-connected discrete surfaces with any number of
boundaries.
In order to derive those correlation functions, a matrix approach of the problem, similar to the one developed in
\cite{eynprats}, is used.

\bd\label{defWmatr}
Let k be a positive integer. Let $\pi$ and $\pi$' be two permutations of $\Sigma_k$ and decompose $\pi'^{-1} \pi$ into the product of its irreducible cycles:
\beq
\pi'^{-1} \pi = P_1 P_2 \dots P_n
\eeq
Each cycle $P_i$ of $\pi'^{-1} \pi$, of length $p_i$, is denoted :
\beq
P_m = (i_{m,1} \rightarrow^{\pi} j_{m,1} \leadsto^{\pi'^{-1}} i_{m,2} \rightarrow^{\pi} j_{m,2}\leadsto^{\pi'^{-1}}  \dots \leadsto^{\pi'^{-1}} i_{m,p_m} \rightarrow^{\pi} j_{m,p_m} \leadsto^{\pi'^{-1}} i_{m,1})
\eeq

For any $(x_1,y_1,x_2,y_2, \dots , x_k,y_k) \in \mathbf{C}$, we define :
\beq
{\cal{W}}^{k}_{\pi,\pi'}(x_1,y_1, \dots , x_k,y_k):= \left\langle\prod_{m=1}^{n} \left( \delta_{p_m,1} + {1\over N}\Tr \prod_{j=1}^{p_m} {1 \over (M_1-x_{i_{m,j}})(M_2-y_{j_{m,j}})}\right) \right\rangle
\eeq
which is a $k!\times k!$ matrix.
\ed

Let us now generalize the notion of planarity of a permutation.

\bd\label{defcoplanarity}
Let k be a positive integer. Let $\pi$ and $\pi$' be two permutations of $\Sigma_k$.

A permutation $\sigma \in \Sym_k$ is said to be planar wrt $(\pi , \pi')$ if
\beq
n_{\rm cycles}(\pi^{-1} \sigma)+n_{\rm cycles}(\pi'^{-1} \sigma)=k+n_{\rm cycles}(\pi'^{-1} \pi)
\eeq

Let $\Sym_k^{(\pi,\pi')} \subset \Sym_k$ be the set of permutations planar wrt $(\pi , \pi')$.

Graphically, if one draws the sets of points $(x_{i_{1,1}},y_{j_{1,1}},x_{i_{1,2}},y_{j_{1,2}}, \dots , x_{i_{1,p_1}},y_{j_{1,p_1}})$,
$(x_{i_{2,1}},y_{j_{2,1}},x_{i_{2,2}},y_{j_{2,2}}, \dots , x_{i_{2,p_2}},y_{j_{2,p_2}})$, $\dots$ , $(x_{i_{p,1}},y_{j_{p,1}},x_{i_{p,2}},y_{j_{p,2}}, \dots , x_{i_{n,p_n}},y_{j_{n,p_n}})$
on n circles and link each pair $(x_j, y_{\sigma(j)})$ by a line, these lines do not intersect nor go from one circle to another.

\ed

\br One can straightforwardly see two properties of these sets:
\begin{itemize}
\item This relation of planarity wrt to $(\pi , \pi')$ is symmetric in $\pi$ and $\pi'$, that is to say:
\beq
\Sym_k^{(\pi,\pi')} = \Sym_k^{(\pi',\pi)}
\eeq
\item The planarity defined in \eq{Defplanar} corresponds to $\pi = Id$ and $\pi'= S^{-1}$:
\beq
\Symp_k = \Sym_k^{(Id,S^{-1})}
\eeq

\end{itemize}
\er

Directly from these definitions and the preceding results comes the following theorem computing any generating function of discrete
surface with boundaries.

\bt\label{thWCsigmamat}
\beq\encadremath{
{\cal{W}}^{k}_{\pi,\pi'}(x_1,y_1,x_2,y_2, \dots , x_k,y_k) = \sum_\sigma {\cal{C}}_{\sigma,\pi,\pi'}^{k}(x_1,y_1, \dots , x_k,y_k) \prod_{i=1}^k W_1(x_i,y_{\sigma(i)})
}\eeq

where ${\cal{C}}_{\sigma}^k$ is the $k! \times k!$-matrix defined by:
\begin{itemize}
\item  ${\cal{C}}_{\sigma,\pi,\pi'}^{k}(x_1,y_1,x_2,y_2, \dots , x_k,y_k) := 0$ if $\sigma$ is not planar wrt $(\pi,\pi')$;

\item if $\sigma$ is planar wrt $(\pi,\pi')$ :
\bea\label{defCm}
&&{\cal{C}}_{\sigma,\pi,\pi'}^k (x_1,y_1, \dots , x_k,y_k) :=\cr
&&\prod_{m=1}^a F^{(a_m)}(x_{r_{m,1}},y_{\sigma(r_{m,1})},x_{r_{m,2}},y_{\sigma(r_{m,2})},\dots,x_{r_{m,a_m}},y_{\sigma(r_{m,a_m})}) \cr
&&\times \prod_{m=1}^\td{a} F^{(\td{a}_m)}(x_{\td{r}_{m,1}},y_{\sigma(\td{r}_{m,1})},x_{\td{r}_{m,2}},y_{\sigma(\td{r}_{m,2})},\dots,x_{\td{r}_{m,a_m}},y_{\sigma(\td{r}_{m,a_m})}) \cr
\eea

with the decompositions of $\pi^{-1} \sigma$ and $\pi'^{-1} \sigma$ into their products of cycles:
\beq
\pi^{-1} \sigma = \pi_1 \pi_2 \dots \pi_a \virg \pi'^{-1} \sigma = \td{\pi}_1 \td{\pi}_2 \dots \td{\pi}_{\td{a}}
\eeq

such that:
\beq
\pi_m = (r_{m,1},r_{m,2}, \dots , r_{m,a_m}) \virg \td{\pi}_m = (\td{r}_{m,1},\td{r}_{m,2}, \dots , \td{r}_{m,\td{a}_m})
\eeq
\end{itemize}

\et

\br
From the definition, one can see that $\sigma(r_{m,a_m}) = \pi(r_m,1)$ and $\sigma(\td{r}_{m,a_m}) = \pi'(\td{r}_m,1)$ for any $m$.
\er

\subsection{Properties of the $C_\sigma^k$'s.}

\bl
The matrices ${\cal{C}}_\sigma^k$ are symmetric.
\el

\proof{ It comes directly from the definition.}

\bl
\beq
\sum_{\sigma} {\cal{C}}_\sigma^k = Id
\eeq
\el

\proof{
One has:
\beq
{\cal{W}}^{k}_{\pi,\pi'}(x_1,y_1,x_2,y_2, \dots , x_k,y_k) = \sum_\sigma {\cal{C}}_{\sigma,\pi,\pi'}^{k} \prod_{i=1}^k W_1(x_i,y_{\sigma(i)})
\eeq
Let us shift all the $x$'s by a translation $a$ and send $a\to\infty$, i.e. replace all the $x_i$'s by $x_i+a$.
In the LHS, only the $\delta$-terms of Definition \ref{defWmatr} survive in the limit $a\to\infty$, and thus the LHS tends towards the identity matrix.
In the RHS, notice that $W_1(x_i+a,y_{\sigma(i)}) \rightarrow 1$.
And  ${\cal{C}}_{\sigma}^{k}$, which depends only on the differences between $x_i$'s, is independent of $a$.
}

\subsection{Some commutation properties}

\bd
Let ${\cal M}^k(\vec{x},\vec{y},\xi,\eta)$ be the $k! \times k!$ matrix defined by:
\beq
{\cal M}^k(\vec{x},\vec{y},\xi,\eta)_{\pi,\pi'}:= \prod_{i} ( \delta_{\pi(i),\pi'(i)} + {1\over (\xi-x_i)(\eta-y_{\pi(i)})})
\eeq

Let ${\cal{A}}^{(k)}(x_1,y_1, \dots, x_k,y_k)$ be the $k! \times k!$ matrix defined by:
\beq
\left\{
\begin{array}{l}
{\cal{A}}^{(k)}_{\pi,\pi}(x_1,y_1, \dots, x_k,y_k) := \sum_i x_i y_{\pi(i)} \cr
{\cal{A}}^{(k)}_{\pi,\pi'}(x_1,y_1, \dots, x_k,y_k) := 1 \; \rm{if} \; \pi \pi'^{-1} = \rm{transposition} \cr
{\cal{A}}^{(k)}_{\pi,\pi'}(x_1,y_1, \dots, x_k,y_k) := 0 \; \rm{otherwise}
\end{array}
\right.
\eeq

\ed

\bt\label{thcommut}
\beq\encadremath{
\forall \sigma,\xi,\eta\, , \qquad [{\cal M}^k(\vec{x},\vec{y},\xi,\eta),{\cal C}_{\sigma}^{k}(\vec{x},\vec{y})]= 0
}\eeq
and
\beq\encadremath{
\forall \xi,\eta\, , \qquad [{\cal M}^k(\vec{x},\vec{y},\xi,\eta),{\cal W}^{k}(\vec{x},\vec{y})]= 0
}\eeq
\et

\proof{
Let us define:
\beq
\widetilde{\cal{M}}(\vec{x},\vec{y},\xi,\eta) := {\cal{M}}(N \vec{x}, \vec{y}, N \xi, \eta)
\eeq
and
\beq
\widetilde{\cal{W}}^{k}_{\pi,\pi'}(x_1,y_1, \dots , x_k,y_k):= \left\langle\prod_{m=1}^{n} \left( \delta_{p_m,1} + \Tr \prod_{j=1}^{p_m} {1\over N}{1 \over (M_1-x_{i_{m,j}})(M_2-y_{j_{m,j}})}\right) \right\rangle
\eeq

It was proven in \cite{eynprats} that:
\beq
 [\widetilde{\cal M}^k(\vec{x},\vec{y},\xi,\eta),\widetilde{\cal W}^{k}(\vec{x},\vec{y})]= 0
\eeq

Now, in the large $N$ limit, the factorization property \cite{ZJDFG} $<\Tr \Tr>\sim <\Tr><\Tr>$, implies:
\bea
&& \widetilde{\cal{W}}^{k}_{\pi,\pi'}(x_1,y_1,x_2,y_2, \dots , x_k,y_k) \cr
&\sim& N^{n_{\rm cycles}(\pi'^{-1} \pi)-k}
\prod_{m=1}^{n} W_{p_m}(x_{i_{m,1}},y_{\pi(i_{m,1})},\dots,x_{i_{m,p_m}},x_{\pi(i_{m,p_m})}) \cr
&\sim& N^{n_{\rm cycles}(\pi'^{-1} \pi)-k} {\cal{W}}^{k}_{\pi,\pi'}(x_1,y_1,x_2,y_2, \dots , x_k,y_k) \cr
\eea
and using theorem \ref{thWCsigmamat}, we have:
\bea
&& \widetilde{\cal{W}}^{k}_{\pi,\pi'}(x_1,y_1,x_2,y_2, \dots , x_k,y_k) \cr
&\sim& N^{n_{\rm cycles}(\pi'^{-1} \pi)-k}
 \sum_\sigma {\cal{C}}_{\sigma,\pi,\pi'}^{k}(x_1,y_1,x_2,y_2, \dots , x_k,y_k) \prod_{i=1}^k W_1(x_i,y_{\sigma(i)})
\eea

Notice that
\bea
&& {\cal{C}}_{\sigma,\pi,\pi'}^{k}(x_1,y_1,x_2,y_2, \dots , x_k,y_k)\cr
&=& N^{k-n_{\rm cycles}(\pi^{-1} \sigma)+k-n_{\rm cycles}(\pi'^{-1} \sigma)}
{\cal{C}}_{\sigma,\pi,\pi'}^{k}(Nx_1,y_1,Nx_2,y_2, \dots , Nx_k,y_k)\cr
&=& N^{k-n_{\rm cycles}(\pi'^{-1} \pi)}
{\cal{C}}_{\sigma,\pi,\pi'}^{k}(Nx_1,y_1,Nx_2,y_2, \dots , Nx_k,y_k)\cr
\eea

Thus:
\bea
&& \widetilde{\cal{W}}^{k}_{\pi,\pi'}(x_1,y_1,x_2,y_2, \dots , x_k,y_k) \cr
&\sim& \sum_\sigma {\cal{C}}_{\sigma,\pi,\pi'}^{k}(Nx_1,y_1,Nx_2,y_2, \dots , Nx_k,y_k) \prod_{i=1}^k W_1(x_i,y_{\sigma(i)})
\eea
Then, from \cite{eynprats}, we have:
\beq
0=\sum_\sigma \left[{\cal{M}}^{k}(N \vec{x}, \vec{y}, N \xi, \eta),  {\cal{C}}_{\sigma}^{k}(Nx_1,y_1, \dots , Nx_k,y_k)  \right]\prod_{i=1}^k W_1(x_i,y_{\sigma(i)})
\eeq
In particular, choose a permutation $\sigma$, and  take the limit where $y_i\to Y(x_{\sigma^{-1}(i)})$,
you get in that limit:
\beq
0= \left[{\cal{M}}^{k}(N \vec{x}, \vec{Y(x_{\sigma^{-1}})}, N \xi, \eta),  {\cal{C}}_{\sigma}^{k}(Nx_1,Y(x_{\sigma^{-1}(1)}), \dots , Nx_k,Y(x_{\sigma^{-1}(k)}))  \right]
\eeq
Since this equation holds for any potentials $V_1$ and $V_2$, it holds for any function $Y(x)$, and thus the $Y(x_i)$'s can be chosen independentely of the $x_i$'s, and thus, for any $y_1,\dots, y_k$,
we have:
\beq
0= \left[{\cal{M}}^{k}(N \vec{x}, \vec{y}, N \xi, \eta),  {\cal{C}}_{\sigma}^{k}(Nx_1,y_1, \dots , Nx_k,y_k)  \right]
\eeq
Since it holds for any $x_i$'s and $\xi$, it also holds for $x_i/N$ and  $\xi/N$.}

\bc\label{corcommut}
\beq
\forall \sigma\, , \qquad [{\cal A}^{(k)}(\vec{x},\vec{y}),{\cal C}_{\sigma}^{k}(\vec{x},\vec{y})]= 0
\eeq
\ec
\proof{
The corollary is obtained by taking the large $\xi$ and $\eta$ limit of theorem \ref{thcommut} (see Appendix of \cite{eynprats}).
}

\subsection{Examples: $k=2$.}

\bea
{\cal{W}}^{(2)} = \pmatrix{W_{11}W_{22} & {W_{11}W_{22}-W_{12}W_{21}\over (x_1-x_2)(y_1-y_2)}\cr
{W_{11}W_{22}-W_{12}W_{21}\over (x_1-x_2)(y_1-y_2)} & W_{12}W_{21} }
\eea

where $W_{ij} = W_{1}(x_i,y_j)$.

\bea
{\cal C}^2_{Id} = \pmatrix{1 & {1\over (x_1-x_2)(y_1-y_2)}\cr {1\over (x_1-x_2)(y_1-y_2)} & 0 }
\eea
\bea
{\cal C}^2_{(12)} = \pmatrix{0 & {1\over (x_1-x_2)(y_2-y_1)}\cr {1\over (x_1-x_2)(y_2-y_1)} & 1 } = 1- {\cal C}^2_{Id} \cr
\eea

\newsection{Application: Gaussian case}

There is an example of special interest, in particular for its applications to string theory in the BMN limit \cite{BMN},
it is the Gaussian-complex matrix model case, $V_1=V_2=0$.
In that case one has $E(x,y)=xy-1$, and thus:
\beq
W_1(x,y)={xy\over xy-1}
\eeq

The loop equation defining recursively the $W_k$'s can be written:
\bea
&& (x_1 y_k-1)W_k(x_1,y_1,\dots,x_k,y_k) = \cr 
&& x_1 \sum_{j} {W_{j-1}(x_j,y_1,\dots,y_{j-1})-W_{j-1}(x_1,y_1,\dots,y_{j-1})\over x_1-x_j}\cr
&& \quad \times W_{k-j+1}(x_j,y_j,\dots,x_k,y_k) \cr
\eea

Its solution is then:
\beq
W_k(x_1,y_1,\dots,x_k,y_k) = \sum_{\sigma\in \Sym_k}\, C^{(k)}_\sigma(x_1,y_1,\dots,x_k,y_k)\,\, \prod_{i=1}^k {x_i y_{\sigma(i)} \over x_i y_{\sigma(i)} -1 }
\eeq

From the loop equation, one can see that $W_k(x_1,y_1,\dots,x_k,y_k)$ may have poles only when $x_i \rightarrow y_j^{-1}$ for any $i$ and $j$.
Because the $C_\sigma$'s are rational functions of all their variables and because $W_k$ has no singularity when $x_i=x_j$ or $y_i=y_j$, one can write:
\beq
W_k(x_1,y_1,\dots,x_k,y_k) = {N_k(x_1,y_1,x_2,y_2, \dots , x_k,y_k) \over \prod_{i,j}(x_i y_j-1)}
\eeq

where  $N_k(x_1,y_1,x_2,y_2, \dots , x_k,y_k)$ is a polynomial in all its variables.

Moreover, the loop equation taken for the values $x_k=0$ or $y_k=0$ shows that $W_k(x_1,y_1,\dots,0,y_k) = W_k(x_1,y_1,\dots,x_k,0)=0$.
Using the cyclicity property of $W_k(x_1,y_1,\dots,x_k,y_k)$, one can claim that it vanishes whenever one of its arguments is equal to 0.
One can thus factorize the polynomial $N_k(x_1,y_1,x_2,y_2, \dots , x_k,y_k)$ as follows:
\beq
W_k(x_1,y_1,\dots,x_k,y_k) = {Q_k(x_1,y_1,x_2,y_2, \dots , x_k,y_k) \prod_{i} x_i y_i \over \prod_{i,j}(x_i y_j-1)}
\eeq
where $Q_k(x_1,y_1,\dots,x_k,y_k)$ is a polynomial of degree $k-2$ with integer coefficient in all its variables.

Notice that $Q_k(x_1,y_1,\dots, y_{\sigma(i)}^{-1},y_i, \dots ,x_k,y_k)=0$ if $\sigma$ is not planar.

As an example, we have:
\begin{itemize}
\item for $k=2$:
\beq
W_2(x_1,y_1,x_2,y_2) = {x_1x_2 y_1 y_2\over \prod_{i,j}(x_i y_j-1)} \virg Q_2(x_1,y_1,x_2,y_2) = 1
\eeq

\item for $k=3$:
\beq
W_3(x_1,y_1,x_2,y_2,x_3,y_3) = (2-\sum_i x_i y_{i+1} + x_1 x_2 x_3 y_1 y_2 y_3)\, {x_1 x_2 x_3 y_1 y_2 y_3\over \prod_{i,j}(x_i y_j-1)}
\eeq
and
\beq
Q_3(x_1,y_1,x_2,y_2,x_3,y_3) = (2- x_1y_2 - x_2 y_3 - x_3 y_1 + x_1 x_2 x_3 y_1 y_2 y_3)
\eeq

\end{itemize}

\newsection{Conclusion}

In this article, we have computed the generating functions of discs with all possible boundary conditions,
i.e. the large $N$ limit of all correlation functions of the formal 2-matrix model.
We have found that the $2k$ point correlation function can be written like the Bethe Ansatz for the $\delta$-interacting bosons,
i.e. a sum over permutations of product of 2-point functions.
That formula is universal, it is independent of the potentials.

An even more powerful approach consists in gathering all possible $2k$ point correlation functions in a $k!\times k!$ matrix ${\cal W}^k$.
We have found that this matrix ${\cal W}^k$ satisfies commutation relations with a family of matrices ${\cal M}^k$ which depend on two spectral parameters,
and are related to the representations of $U(n)$ \cite{eynprats}.
We claim that the theorem \ref{thcommut} is almost equivalent to the loop equations, and allows to determine ${\cal W}^k$.

\medskip

It remains to understand how all these matrices and coefficients $C_\sigma$ are related to usual formulations of integrability,
i.e. how to write these in terms of Yang Baxter equations. For instance, the similarity with equations found
in Razumov-Stroganov conjecture's proof \cite{RazStro} is to be understood.

\medskip

One could also hope to find a direct proof of theorem \ref{mainth}, without having to solve the loop equations.
In other words, we have found that the $2k$-point function can be written only in terms of $W_1$, while, in the derivation, we use the one point functions $Y(x)$ and $X(y)$ although they don't appear in the final result.

\medskip

The next step, is to be  able to compute the $1/N^2$ expansion of those correlation functions, as well as the large $N$ limit of
connected correlation functions.
We are already working that out, by mixing the approach presented in the present article and the Feynman graph approach of \cite{eynloop1mat}
generalized to the 2-matrix model in \cite{eoloop2mat}.

\medskip

Another prospect is to go to the critical limit, i.e. where we describe generating functions for continuous surfaces with conformal invariance,
and interpret this as boundary conformal field theory \cite{kostov}.

\subsection*{Acknowledgements}
The authors want to thank  M. Bauer, M. Berg\`ere, F. David, P. Di Francesco, J.B. Zuber for stimulating discussions.
This work was partly supported by the european network Enigma (MRTN-CT-2004-5652).

\eop

%%%%%%%%%%%%%%%%%%%%%%%%%%%%%%%%%%%%%%%%%%%%%%%%%
%%%%%%%%%APPENDIX
\setcounter{section}{0}
\appendix{Practical computation of $f_\sigma$ }
\label{app}
In this appendix, we build a set of trees in bijection with $\Symp_k$ and use it in order to compute practicaly
the weights $f_\sigma$ defined in Definition \ref{deff}.

\bd
Let ${\cal T}_k$ be the set of trees defined as follows:
A tree $T$ belongs to $T\in {\cal T}_k$, and is called a $k$-planar tree if and only if:
\begin{itemize}
\item its root is labelled $k+1$;
\item it is composed of k+1 vertices, labeled by $[1, \dots, k+1]$;
\item it has $k$ edges which can be either upgoing or downgoing;
\item its vertices have valence $1,2$ or $3$, and are of one of the following eight possibilities, in which the point $m$ denotes the origin of the branch containing $i$:
\begin{itemize}
\item two trivalent vertices :
\beq
\begin{array}{r}
{\epsfysize 1.5cm\epsffile{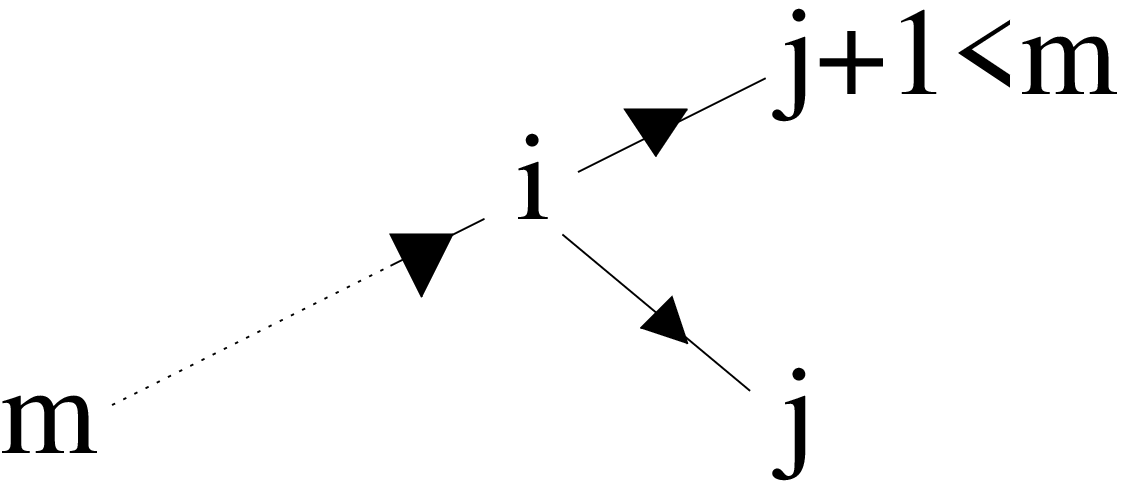}}
\end{array}
\and
\begin{array}{r}
{\epsfysize 1.5cm\epsffile{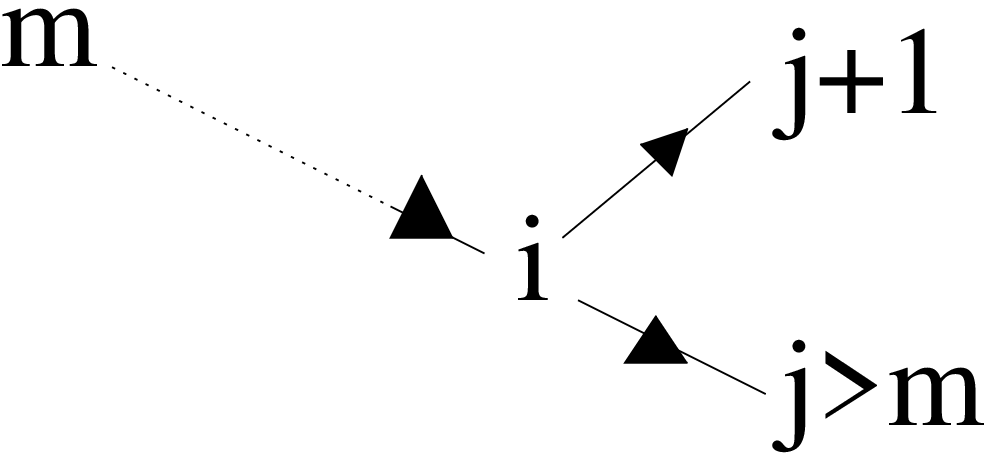}}
\end{array}
\eeq

\item four bivalent vertices :
\beq
\begin{array}{r}
{\epsfysize 1.4cm\epsffile{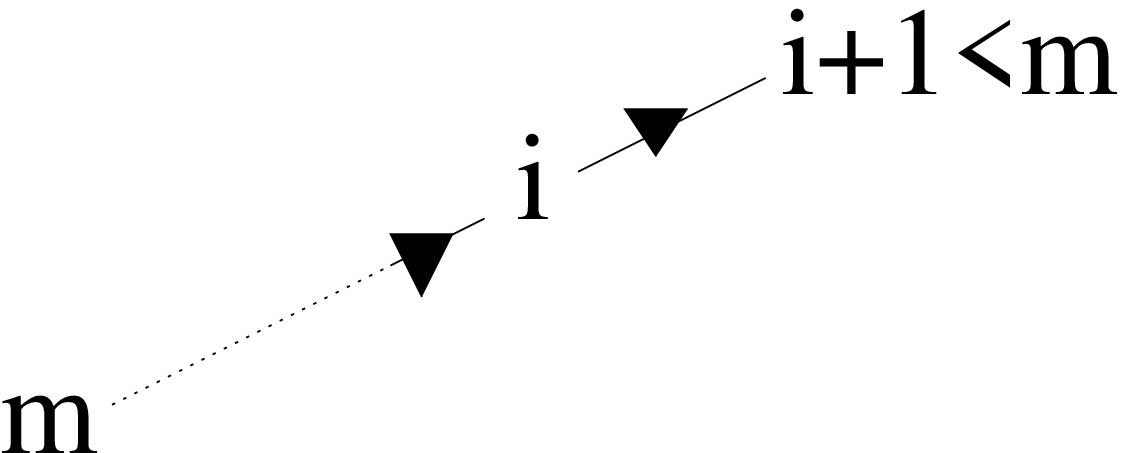}}
\end{array}
\virg
\begin{array}{r}
{\epsfysize 1.4cm\epsffile{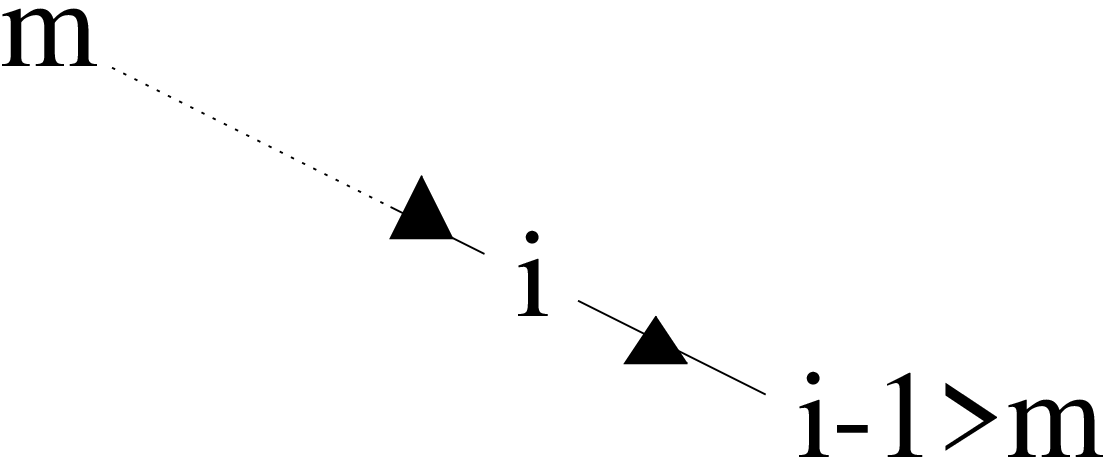}}
\end{array}
\eeq

\beq
\begin{array}{r}
{\epsfysize 1cm\epsffile{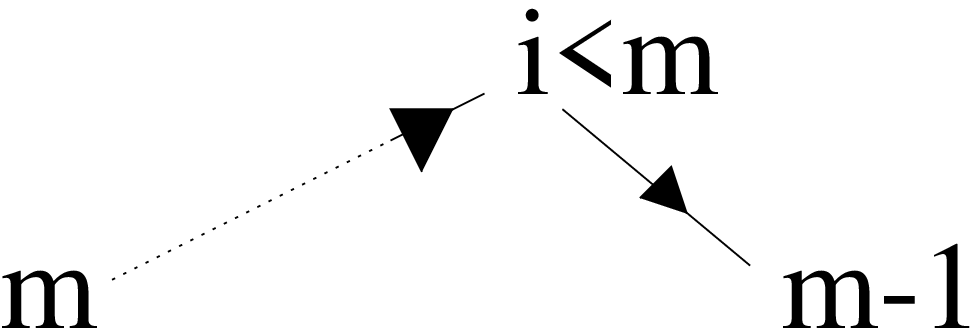}}
\end{array}
\and
\begin{array}{r}
{\epsfysize 1cm\epsffile{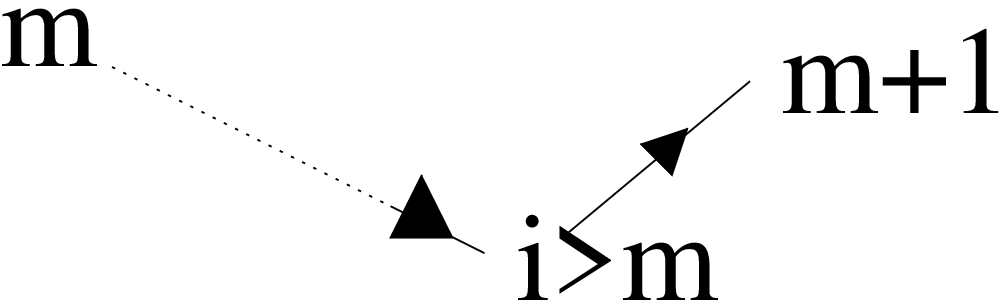}}
\end{array}
\eeq

\item two monovalent vertices corresponding to the leaves of the tree:
\beq
\begin{array}{r}
{\epsfysize 1cm\epsffile{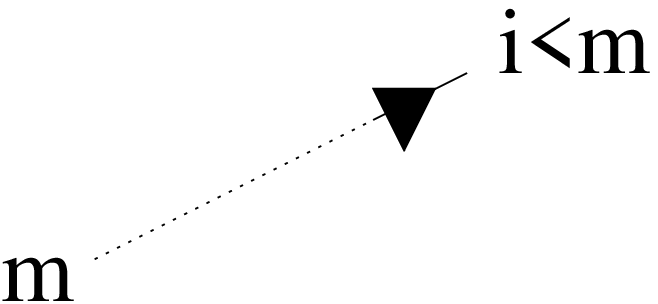}}
\end{array}
\and
\begin{array}{r}
{\epsfysize 1cm\epsffile{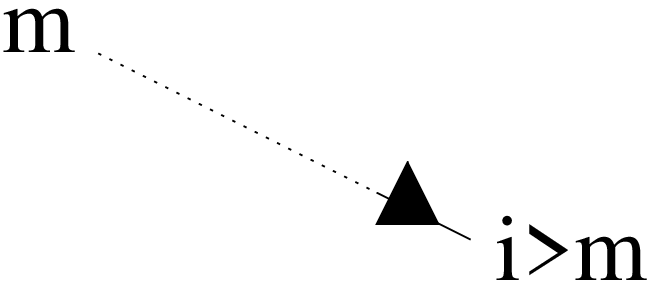}}
\end{array}
\eeq

\end{itemize}

\end{itemize}
\ed

\br
Those trees are often called planar binary skeleton trees.
\er

\br
One can see that the first edge is necessarily upgoing, and its extremity is necessarily $1$.
\beq
\begin{array}{r}
{\epsfysize 1.5cm\epsffile{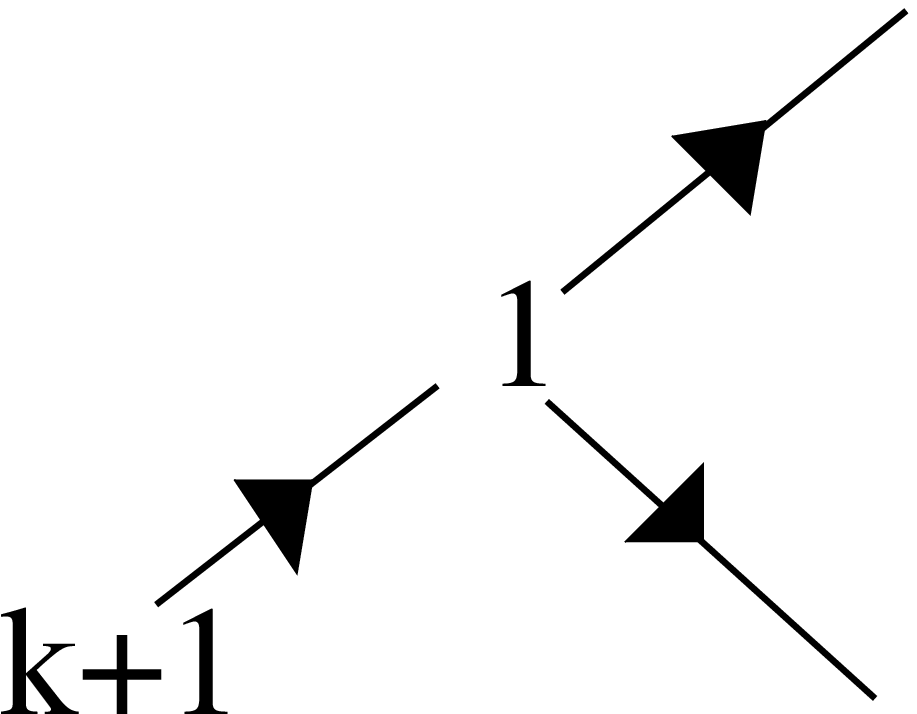}}
\end{array}
\eeq

\er

\bt
There is a bijection between ${\cal T}_k$ and the set of planar permutations $\Symp_{k}$:
\et

\proof{
We build explicitly this bijection between $\Symp_{k-1}$ and ${\cal T}_{k-1}$.
\vs

Consider a planar permutation $\sigma \in \Symp_{k-1}$.
Planarity means that $\sigma$ defines a partition of the disc into faces of two kinds.
Let us say that faces which correspond to cycles of $S\sigma$ are colored in white,
faces which correspond to cycles of $\sigma$ are colored in black.

Decompose $\sigma$ and $S\sigma$ into products of irreducible cycles:
\beq
\sigma=\sigma_1 \sigma_2 \dots \sigma_l
\virg
S\sigma=\td\sigma_1 \td\sigma_2 \dots \td\sigma_{\td{l}}
\eeq
and we assume that $\td\sigma_1$ and $\sigma_1$ contain $x_1$.

Because of planarity, we can define a distance of faces (i.e. cycles) to the face $\td\sigma_1$,
as the number of edges one has to cross for going from a face $\sigma_i$ or $\td\sigma_i$ to $\sigma_1$, and call it $D(\sigma_i)$ or $D(\td\sigma_i)$.

We also define the ``origin'' of a face, noted $m(\sigma_i)$ or $m(\td\sigma_i)$, as follows:
If the face is $\td\sigma_1$, we define $m(\td\sigma_1)=k$, otherwise,
because of planarity, there is only one neighbouring face which is at smaller distance of $\td\sigma_1$.
Because of planarity, those two faces share at most one $x$, and the origin is defined as the label of that $x$.

Thus, each face has a color, white or black, a distance $D$, and an origin $m$.

\smallskip

Now, to every face we associate a branch as follows:

{\noindent $\bullet$} to a white face, $\td\sigma_j$, i.e. a cycle of $S\sigma$, noted
\beq
\td\sigma_j = (\td{i}_{j,1},\td{i}_{j,2},\dots,\td{i}_{j,\td{l}_j})
\virg \td{i}_{j,1}=m(\td\sigma_j) \virg \sigma(\td{i}_{j,n})=\td{i}_{j,n+1}-1
\eeq
we associate the upgoing branch $\td{i}_{j,1}\to\td{i}_{j,2}\to\dots\to\td{i}_{j,\td{l}_j}$
\beq
\begin{array}{r}
{\epsfysize 2.5cm\epsffile{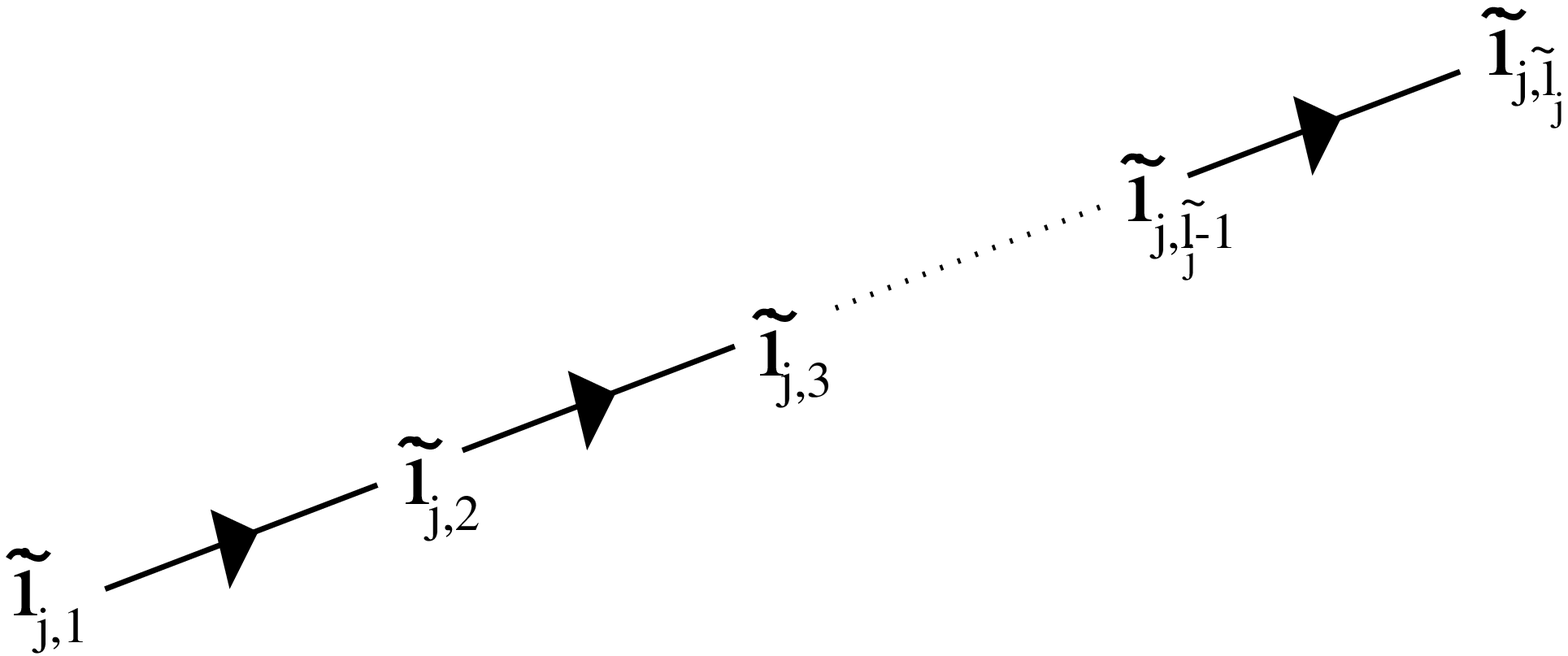}}
\end{array}
\eeq
(if $\td{l}_j=1$, the sequence contains only one vertex $\td{i}_{j,1}=m(\td\sigma_j)$ and no edge).

{\noindent $\bullet$} to a black face, $\sigma_j$, i.e. a cycle of $\sigma$, noted
\beq
\sigma_j = ({i}_{j,1},{i}_{j,2},\dots,{i}_{j,{l}_j})
\virg {i}_{j,1}=m(\sigma_j) \virg \sigma({i}_{j,n})={i}_{j,n+1}
\eeq
we associate the downgoing branch ${i}_{j,1}\to {i}_{j,2}\to\dots\to {i}_{j,{l}_j}$
\beq\begin{array}{r}
{\epsfysize 1.5cm\epsffile{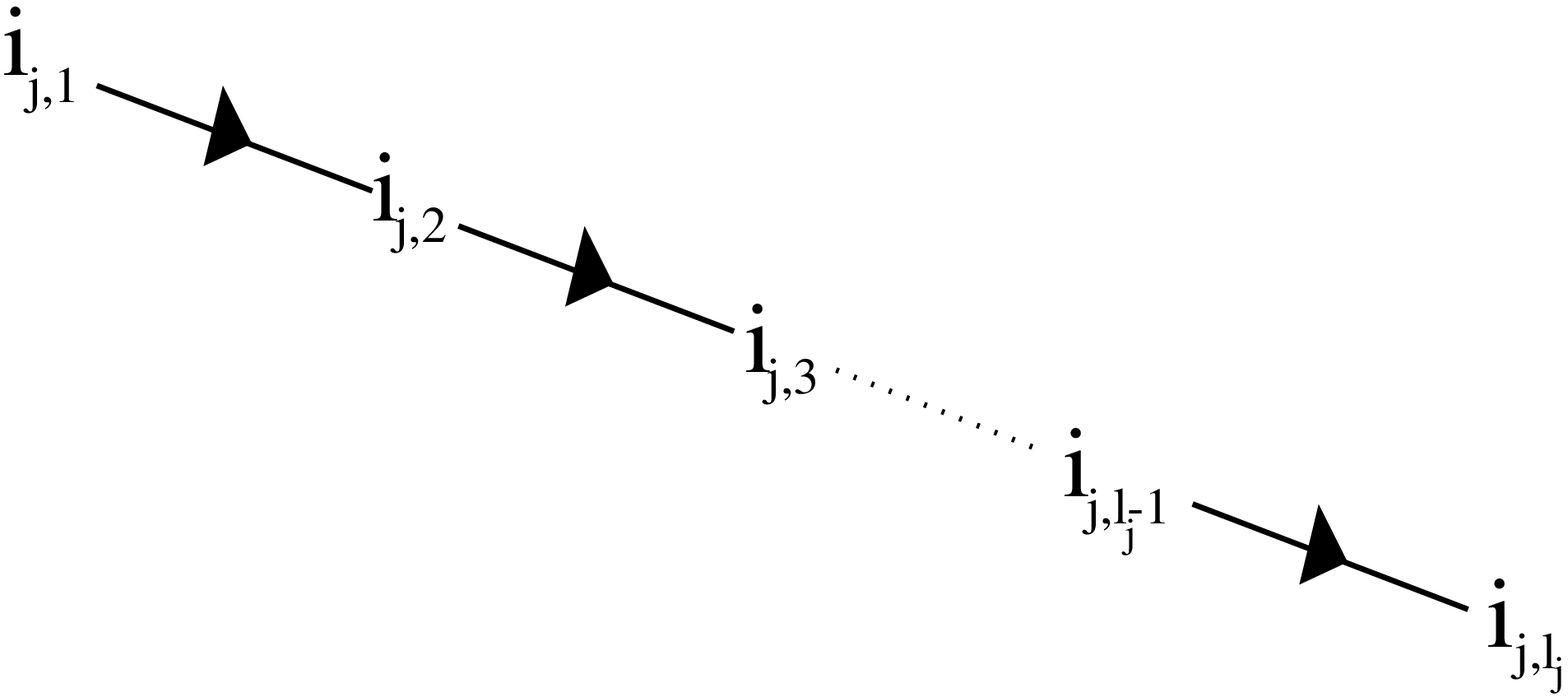}}
\end{array}\eeq
(if ${l}_j=1$, the sequence contains only one vertex ${i}_{j,1}=m(\sigma_j)$ and no edge).

{\noindent $\bullet$} to the first face $\td\sigma_1$,
\beq
\td\sigma_1 = (\td{i}_{1,1},\td{i}_{1,2},\dots,\td{i}_{1,\td{l}_1})
\virg \td{i}_{1,1}=1 \virg \sigma(\td{i}_{1,n})=\td{i}_{1,n+1}-1
\eeq
we associate the upgoing branch $k\to \td{i}_{1,1}\to\td{i}_{1,2}\to\dots\to\td{i}_{1,\td{l}_1}$
\beq\begin{array}{r}
{\epsfysize 1.5cm\epsffile{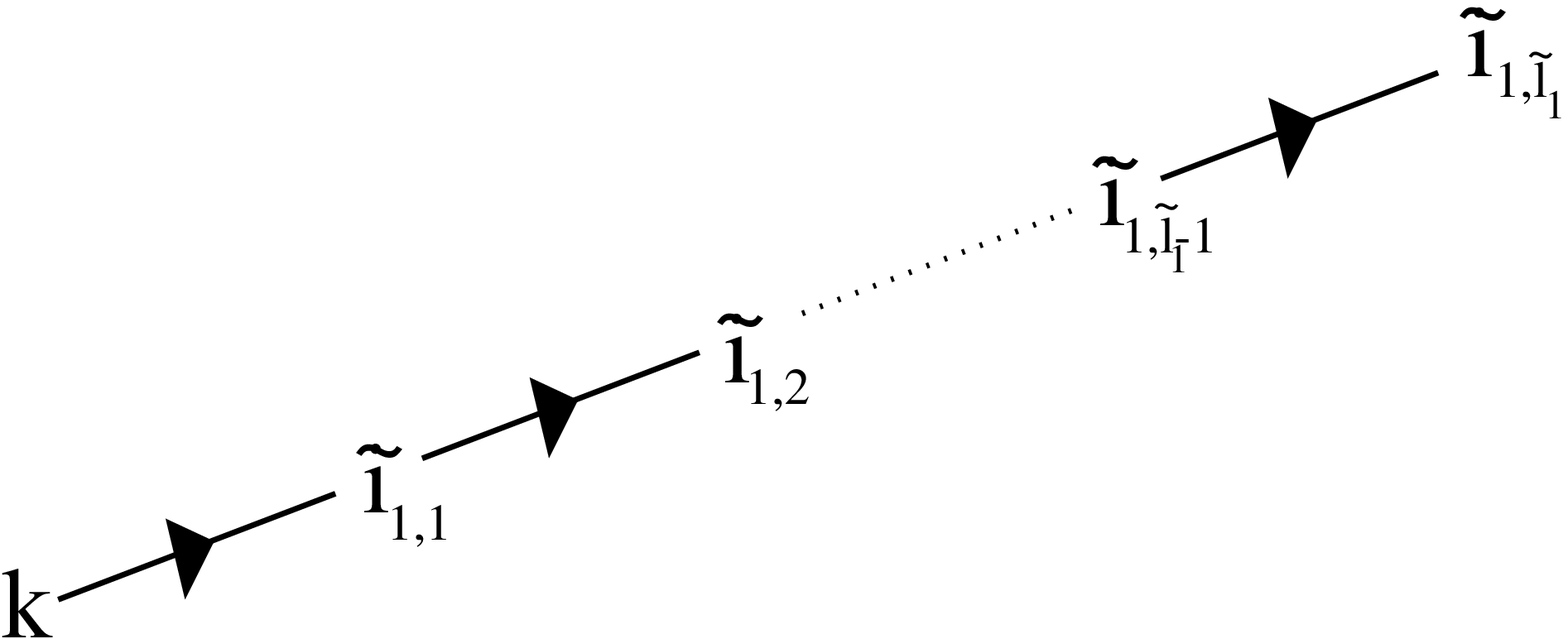}}
\end{array}\eeq

\medskip

By definition of the origin $m$ of a face at distance $D$, the origin of a branch is necessarily a vertex on a branch at distance $D-1$,
and from planarity, it cannot be a vertex on any other branch.
Thus, there is a unique way to attach all branches to their origin, and we obtain a tree, which belongs to ${\cal T}_{k-1}$.

\medskip

\underline{Inverse bijection:}

On the other hand, let us consider a $k-1$-tree. One can build a permutation $\sigma \in \Symp_{k-1}$ as follows:
the image of an element of $(1,\dots , k-1)$ is :
\begin{itemize}
\item its descendant along a downgoing propagator if it exists;
\item the origin of the downgoing branch to which it belongs in the other cases.
\end{itemize}

Because of the form of the vertices, the upgoing branches are necessarily the cycles of $S\sigma$.
And since the branches form a tree, it implies that two faces touch one another through zero or one edge. Thus the permutation $\sigma$ is planar.

 It is easy to see that this application is the inverse of the preceding one.}

{\bf Example:}

Let us carry out explicitly step by step this building for the permutation $\sigma \in \Symp_{12}$ introduced earlier.
Notice that it is enrooted in $12+1=13$.

\beq
\begin{array}{c}
{\epsfxsize 14cm\epsffile{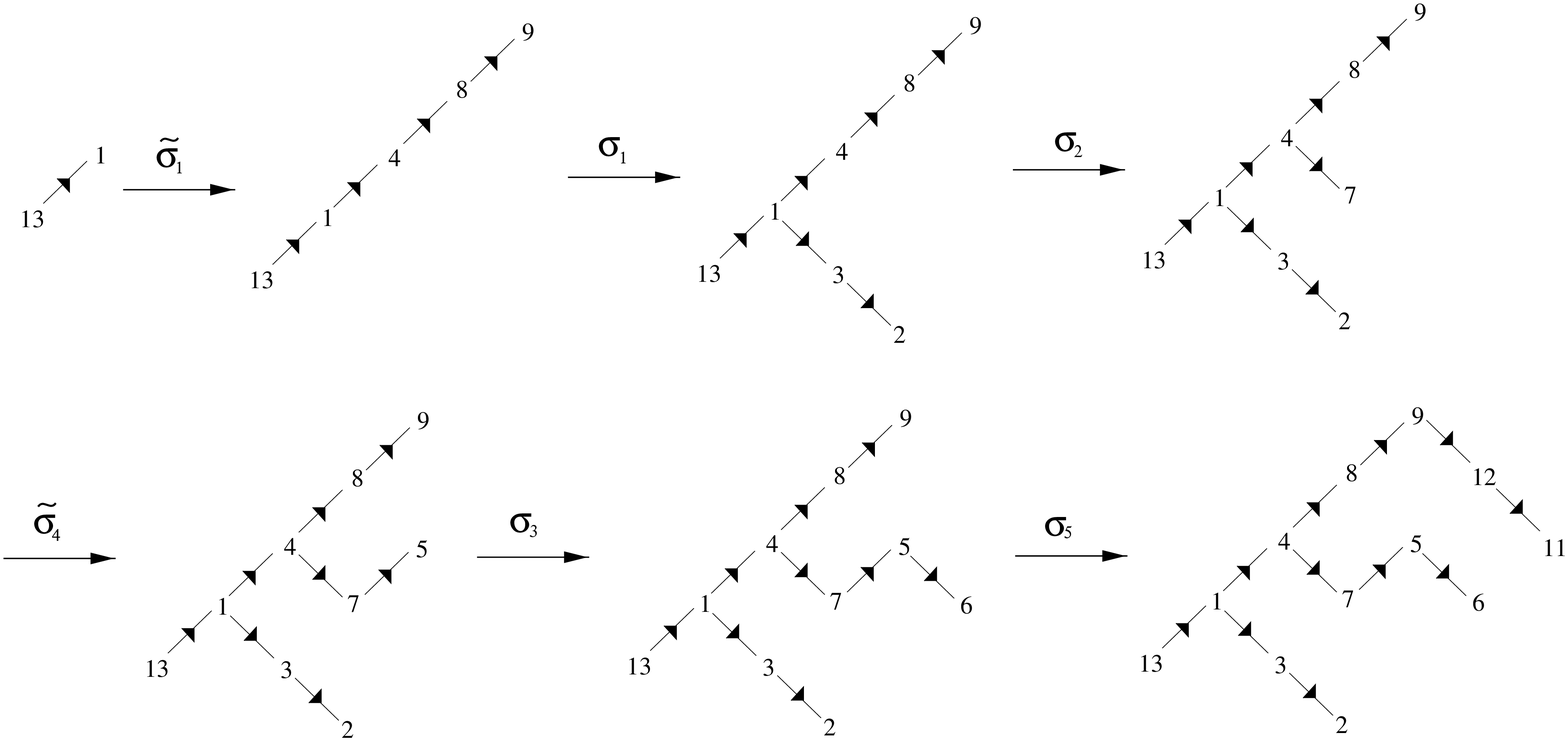}}\cr
\end{array}
\eeq

Considering the last non trivial cycle $\td{\sigma}_6=(10,11)$, one obtains finally the tree corresponding to (\ref{system}):
\beq
\begin{array}{c}
{\epsfxsize 5cm\epsffile{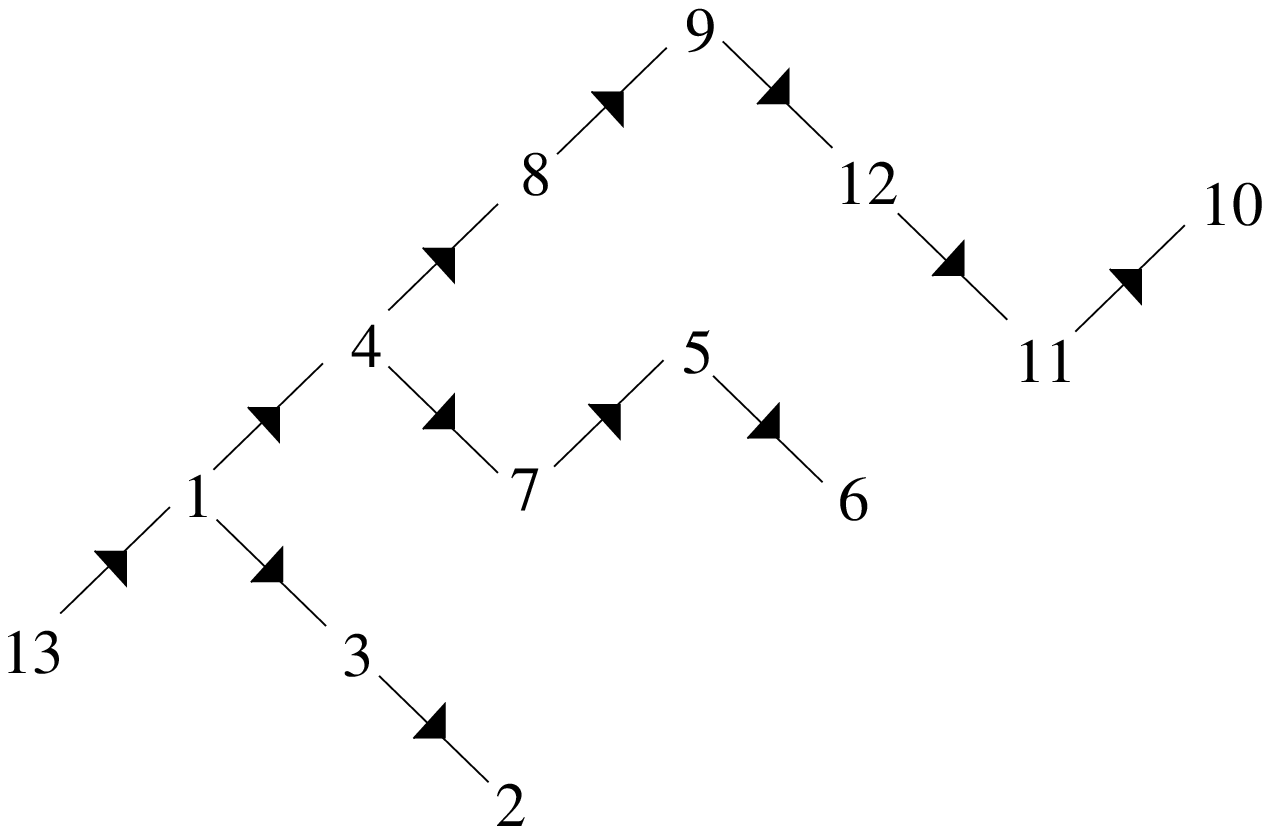}}
\end{array}
\eeq

\bc
$\#{\cal T}_k=\Cat(k)$ , where $\Cat(k)$ is the k'th Catalan number.

\ec

From now, one can see how to simply compute the weight $f_\sigma$ associated to a permutation $\sigma$.

Consider a planar permutation $\sigma \in \Symp_{k}$ and its representation under the form of a tree $T \in \#{\cal T}_k$.
Associate a weight to every edge of the tree as follows:
\begin{itemize}
\item To every downgoing edge $
\begin{array}{r}
{\epsfysize 1cm\epsffile{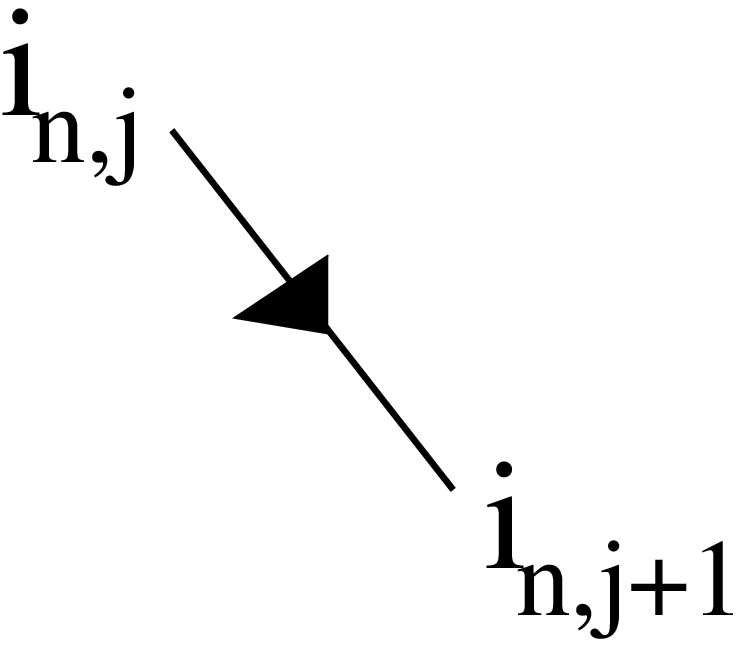}}
\end{array}$ of a cycle $\sigma_n$ of $\sigma$,
one associates the weight $g_{i_{n,1},i_{n,j+1},i_{n,j+2}}$;

\item To every upgoing edge $
\begin{array}{r}
{\epsfysize 1cm\epsffile{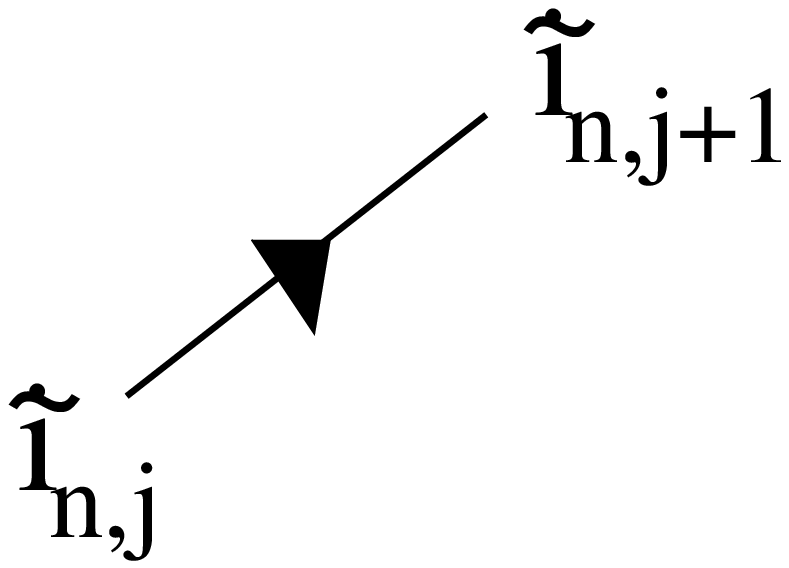}}
\end{array}$ of a cycle $\td{\sigma}_n$  of $S\sigma$,
one associates the weight
$g_{\td{i}_{n,j+1},\td{i}_{n,1},\sigma (\td{i}_{n,j+1})}$.
\end{itemize}

Then, the $f_\sigma$ is the product of all the weights of edges composing $T$.

%%%%%%%%%%%%%%%%%%%%%%%%%%  Bibliography
%%%%%%%%%%%%%%%%%%%%%%%%%%%%%%%%%%

\end{document}